%% file: main.tex
\RequirePackage{lineno}
\documentclass[tightenlines,superscriptaddress,10pt,aps,prd]{revtex4-1}
\pdfoutput=1

\usepackage{placeins}
\usepackage{multirow}
\usepackage[utf8]{inputenc}
\usepackage{color}
\usepackage{rotating}

\usepackage{longtable}
\usepackage{amsmath}
\usepackage{array}
\newcolumntype{L}[1]{>{\raggedright\let\newline\\\arraybackslash\hspace{0pt}}p{#1}}
\newcolumntype{C}[1]{>{\centering\let\newline\\\arraybackslash\hspace{0pt}}p{#1}}
\newcolumntype{R}[1]{>{\raggedleft\let\newline\\\arraybackslash\hspace{0pt}}p{#1}}
\input{common/preamble.tex}

\def\bracketbar{\hbox{\kern-9pt\raise1pt%
    \hbox{{\tiny(}{\lower1.5pt\hbox{\bf--}}{\tiny)}}}}

\begin{document}

\title{Snowmass Neutrino Frontier Report}
\date{\today}
%\date{8-18-2022}
\input{sections/author_list}

\maketitle
%\linenumbers
%\begin{abstract}
%Report abstract goes here.
%\end{abstract}

\input{common/init}

%Use "nfword{}" command for words that should have glossary entries, eg: \nfword{nf}. You can define your own words, abbreviations, and definitions in common/glossary. These definitions will all be printed at the end of the document. This also handles first time using a term and defining it vs using it later without the definition, so use the nfword command any time you use one of these words in the document.

%Use "fixme{}" command to flag things that need work - eg: \fixme{Remove these instructions and write a real summary}

%There are a bunch of standard oscillation parameters, eg: \sinst{13} in common/defs and units, eg: \SI{1300}{km}, in common/units. Use these and add your own as needed.

%Please use standard bibtex entries available from inspire for citations. Enter these in common/citations.bib. For example, the DUNE sensitivity paper \cite{Abi:2020qib} blah blah.

%Plots go in eg: graphics/NF01/. graphics is in the path, so you should not have to type that part. I've made a folder for each of the topical groups to make it easy for each group to curate their own plots, as the document structure may change.

\newpage

\include{execsummary}

\newpage

\section{Introduction}

This document represents a distillation of content generated by the extensive neutrino community activities since the beginning of the Snowmass process in spring of 2020.  

The Neutrino Frontier is organized into ten Topical Groups, listed below.

\begin{itemize}
    \item NF01: \textbf{Neutrino Oscillations.}  Physics of neutrino oscillations in the three-flavor paradigm.
    \item NF02: \textbf{Understanding Experimental Neutrino Anomalies.} 
    Sterile neutrino oscillations, including experiments or theoretical models on 3+1 and 3+N scenarios, as well as other conventional and BSM physics that can explain current short-baseline anomalies or alter disappearance/appearance neutrino oscillation probabilities.
    \item  NF03: \textbf{Beyond the Standard Model.} Searches for signals from physics Beyond the Standard Model (BSM), both from a theoretical and experimental perspective.
    \item NF04: \textbf{Neutrinos from Natural Sources.} Neutrino detection from all natural sources, including the Sun, the Earth, and astrophysical sources. 
    \item NF05: \textbf{Neutrino Properties.} ``Anything about a neutrino," but especially: direct neutrino mass measurements, the Majorana-vs-Dirac nature of the neutrino, and electromagnetic properties of neutrinos.  
    \item NF06: \textbf{Neutrino Interactions.} Neutrino interactions on a wide range of target nuclei and across the full spectrum of neutrino energies. 
    \item NF07: \textbf{Applications.} Potential practical applications of neutrinos, including nuclear non-proliferation, geology, nuclear data, and applications of technology.
    \item NF08/TF11: \textbf{Theory of Neutrino Physics.}   Theoretical aspects of neutrino physics. This is also a topical group within the Theory Frontier.
    \item NF09: \textbf{Artificial Neutrino Sources.} Development, characterization and understanding of human-made neutrino sources, including beams, reactors, and radioactive sources.
    \item NF10: \textbf{Neutrino Detectors.}  Detector technologies capable of exploring neutrino physics across the full spectrum of possible energies, from sub-eV to EeV.

\end{itemize}

More detailed information relevant to each Topical Group area can be found in the respective reports.

The topics considered within these topical groups overlap between groups to a very large extent.  Furthermore, there are significant overlaps with \textit{all} of the other Snowmass Frontiers, perhaps most notably with the Cosmic Frontier, the Theory Frontier and the Instrumentation Frontier.

The content in this report draws heavily on executive summaries of the ten Neutrino Frontier topical group reports. However, we are not organizing the material in this report exactly along the lines of the topical group material, given the many overlaps between topics.  Citations for experimental programs are for the most part omitted, with the exception of some references to Snowmass contributions (``white papers") submitted to the Neutrino Frontier. However a full listing of experiments, including references, provided by the community, can be found in Tab.~\ref{tab:experiments}.

\subsection{Neutrino Frontier Activities}

Below is a brief summary of the community process within the Neutrino Frontier.  Materials can be found at \texttt{https://indico.fnal.gov/category/1101/}.

\begin{itemize}
    \item April 2020: Topical groups formed.
    \item July 2020: Neutrino Town Hall session for initial community input.  Input from this meeting resulted in the formation of a TF11 topical group to represent neutrino physics within the Theory Frontier.  The new group retained an identity within the Neutrino Frontier as NF08.
    \item August 2020: 311 Snowmass Letters of Interest received by the Neutrino Frontier.
    \item Fall 2020: several topical workshops held (some continuing after the pause).
    \item First half of 2021: Snowmass pause.  
    \item Fall 2021: ``Neutrino Frontier White Paper Coordination Workshops." Mini-workshops designed for community members to present and discuss the content of their contributions held.
    \item Jan-Feb 2022: Topical group feedback meetings held to get community input on topical group report drafts.
    \item March 2022 Hybrid Neutrino Frontier meeting at Oak Ridge National Laboratory held.  Neutrino Frontier report executive summary content developed.
    \item March-April 2022: Snowmass Neutrino Colloquia: four sessions of three talks each held to communicate main neutrino concepts and issues to the broader Snowmass community.
    \item July 2022: Community Summer Study in Seattle.
    
\end{itemize}

Overall 86 Snowmass contributions (``white papers") were submitted to the Neutrino Frontier.  Report drafts were opened for several rounds of feedback.  We benefited greatly from the consistent and deep engagement of the community in this process.

\section{Physics Topics}

In this section, we will summarize the main physics questions within the
Neutrino Frontier and strategies for addressing them.  

\subsection{Theory and Motivation}

The discovery of nonzero neutrino masses requires new fundamental fields and new interactions. We know very little about this new physics other than the fact that it exists. The new degrees of freedom can be fermions or bosons, ultra-light or super-heavy, charged or neutral, within reach of experimental efforts being pursued today or virtually invisible to any foreseeable future experiment. At the same time, the discovery of a new mixing matrix -- very different from the quark one -- serves as a new, perhaps decisive, piece to the elusive flavor puzzle.

A robust neutrino theory and phenomenology effort is required in order to exploit the unique probes of fundamental physics offered by neutrino experiments -- to interpret the data, build models to accommodate new phenomena, provide guidance for future experimental efforts, and connect the new discoveries in neutrino physics to other areas of particle and nuclear physics, astrophysics, and cosmology. Neutrino theory requires a broad set of tools in order to attack a unique set of physics problems. Here we highlight the role and the goals of neutrino theory, concentrating on how it complements and contributes to theoretical efforts in other areas of fundamental physics, and on some of the challenges for the near future and the coming decades. 

Among the goals of neutrino theory is to identify the different hypothetical degrees of freedom and interactions responsible for nonzero neutrino masses. More progress requires a coherent theoretical and phenomenological effort to establish connections to other outstanding questions in fundamental particle physics ranging from quantum gravity to the mechanism of baryogenesis to the dark matter puzzle. On the phenomenology side, the physics behind nonzero neutrino masses can manifest itself in, to name a few, neutrino oscillations, fundamental electric-dipole moments, the $g-2$ of charged fermions, charged-lepton flavor violating processes, or high energy colliders. Theory is required in order to explore these connections and identify promising new directions.   

The flavor puzzle -- understanding the underlying fundamental physics that is responsible for the patterns observed in the quark and lepton masses and mixing parameters -- is also the subject of theoretical physics research. It may prove to be one way in which ingredients of a more fundamental theory of nature, including string theory, manifests itself, and it may contain information associated with grand unification. 
%On the other hand, 
The fact we do not yet have all the pieces of the lepton mixing matrix in place allows one to test different general principles that may lurk behind lepton masses and mixing. Expectations from theories of flavor help provide guidance regarding how well we should measure mixing and other fundamental parameters. 

In the three-flavor picture,  all three mixing angles and two mass-squared differences have been measured, and improved precision is required for further progress. In order to fully exploit data from long-baseline oscillation experiments, our quantitative understanding of neutrino scattering cross sections must improve significantly. Furthermore, since neutrinos only interact weakly and the values of  neutrino oscillation parameters are such that for terrestrial experimental setups one is forced to work with neutrino energies below a few GeV,
one needs to understand the scattering cross section of neutrinos off complex nuclei, including carbon, oxygen, and argon. Theoretical nuclear physics is required in order to properly describe the targets and the medium through which the products of the collision propagate. A precise description of relatively low-energy neutrino--nucleon scattering depends on the tools of lattice gauge theory. At lower energies, the physics of coherent elastic neutrino-nucleus scattering, first observed a few years ago, needs input from theoretical nuclear physics. On the other end of the energy spectrum, in order to fully exploit the physics of ultra-high-energy cosmic neutrinos, a solid theoretical understanding of neutrino deep-inelastic scattering is required.  

On the more phenomenological side, neutrino theorists explore, very broadly, the physics potential of long-baseline and short-baseline neutrino oscillation experiments. Phenomenological work has revealed, in the last few years, that experimental setups aimed at precisely measuring oscillations can be used to look for relatively light, new degrees of freedom, including candidates for the dark matter and new fermions that may have something to do with nonzero neutrino masses. This type of work serves as motivation for exploring different options for near-detector complexes and informs decisions regarding promising new detector technologies. Phenomenological and model-building efforts are also required in order to interpret neutrino oscillation data, including current and any future anomalies. In particle physics experiments, in order to fully exploit experimental data, a coherent experimental-phenomenological-model-building effort is required. Neutrino experiments in general, and neutrino oscillation experiments in particular, are no exception. 
 
Searches for the violation of lepton number inform fundamental neutrino physics and are motivated by neutrino theory in different ways. Different theoretical efforts are required in order to connect lepton-number-violating observables to neutrino properties and other new physics, and in order to connect different lepton-number-violating phenomena to one another. The deepest probe of the violation of lepton number is the search for neutrinoless double-beta decay ($0\nu\beta\beta$). The relation between the lifetime of $0\nu\beta\beta$ and fundamental physics parameters requires precise inputs from theoretical nuclear physics and nucleon physics. Current efforts combine state-of-the-art tools from lattice gauge theory, capable of estimating nucleon-level processes, with those of nuclear physics, that define the nontrivial state of these nucleons inside of the complex nuclei of interest.

Neutrinos are also produced in intense astrophysics environments and, together with photons and gravitational waves, serve as messengers between us and the cosmos. Thanks to the weak-interaction nature of neutrino scattering and the very long neutrino lifetime, these neutrinos can and have been detected on Earth, allowing, with nontrivial input from theory, one to learn more about the different astrophysical processes and the properties of the neutrinos themselves. The detection of neutrinos from core-collapse supernova explosions, in particular, provides invaluable information about these violent phenomena, but these data require significant input from theory in physics and astrophysics to interpret.
Incorporating neutrino-flavor information requires understanding the very challenging physics of flavor-transport inside the explosion. 

At higher energies, neutrinos from the cosmos have been detected with energies between 1~TeV and 10~PeV. These observations are expected to shed light, with necessary help from astrophysics and astro-particle theory, on the cosmic ray puzzle and the ultimate particle accelerators in the universe. Ultra-high-energy neutrinos are also unique probes of new phenomena and have been explored extensively by the phenomenology community. Their very long travel distances and extreme energies are sensitive to neutrino properties, including the neutrino lifetime, and, assuming one can distinguish neutrino flavors, allow for powerful tests of the Lorentz invariance and the unitary evolution of quantum mechanical states. 

At the opposite end of the energy spectrum lie the cosmic neutrinos produced at the Big Bang. These have never been directly observed but influence the evolution of the Universe, including the formation of structure. Their existence has been robustly inferred through a variety of different probes, including measurements of the primordial abundances of light nuclei (deuterium, helium, lithium) and precision measurements of the properties of the cosmic microwave background. Theory and phenomenology are required in order to compute the impact of primordial neutrinos and extract, from diverse measurements of the large-scale structure of the Universe, neutrino properties. Some of these translate into nontrivial theoretical and computational physics problems involving many-body physics, nonlinear dynamics, etc. New neutrino properties may be responsible for some of the observed discrepancies in cosmological data, including the $H_0$ puzzle. The exploration of these new properties -- how they impact cosmic surveys, how they can be constrained by Earth-bound experiments -- is the job of particle and astroparticle theorists and cosmologists. 
  
In the last decade, the neutrino theory effort has grown both in the United States and in the rest of the world, but so has the need for a more robust domestic theoretical physics community. Some positive developments include a few dedicated efforts from the Department of Energy and the National Science Foundation, increased investment in neutrino theory in the national laboratories, and a more robust ``neutrino'' footprint in the lattice~\cite{USQCD:2022mmc} and nuclear physics communities. These are necessary but not sufficient; the domestic neutrino theory community is still very small. A dedicated and coherent effort, with significant investment from the funding agencies, enthusiastic commitment and leadership from the present neutrino theory community, and the support of the wider particle theory and neutrino experimental communities is absolutely necessary. To meet the proposed experimental schedules and ambitions, such an increase in the neutrino theory effort is needed now.

\subsection{Three-Flavor Neutrino Oscillation}

\subsubsection{Goals in Three-Flavor Oscillations}

Understanding the flavor structure of particle physics, 
why the fermion masses are distributed as they are and why their mixing 
with the weak interaction has the structure it has, represents one of the biggest gaps in understanding of our model of particle physics. 
Neutrino oscillation measurements of the remaining unknowns, the 
neutrino mass ordering, the octant of $\theta_{23}$, and the value of \deltacp, 
will enhance our knowledge of 
the three-family structure of particle physics. Furthermore, 
precise measurements of the three-flavor oscillation 
parameters will allow consistency tests of the three-flavor 
paradigm which could lead to discovery of additional new physics.

Thanks to the many oscillation experiments over the last several decades, 
a clear picture of the overall framework of three-flavor 
oscillations has emerged. We have two mass-squared differences: \dm{21} and \dm{31}, the magnitudes of which are well 
measured. We know two of the mixing angles 
-- $\theta_{13}$ and $\theta_{12}$ -- fairly well. 
The measured value of the third mixing angle -- $\theta_{23}$ -- 
is consistent with maximal, but with insufficient precision to 
determine its octant if non-maximal, and the complex 
phase \deltacp is largely unconstrained.
The three primary goals of the coming years of 
neutrino oscillation measurements are:
\begin{enumerate}
    \item to determine the ``neutrino mass ordering," which is defined by the sign of \dm{31}, 
    \item to determine whether $\theta_{23}$ is more than or less than 45$\degs$ and how close to maximal it is, known as the ``octant" question, and 
    \item to measure \deltacp and determine if $\sin\deltacp = 0$ can be excluded or not.
\end{enumerate}
Some of these parameters can be assessed at an individual 
experiment (with input from orthogonal experiments), and 
some parameters will first be measured as a result of 
global analyses of all relevant neutrino data, known as global fits. 
Tau neutrino appearance~\cite{Abraham:2022jse} can further over-constrain the picture of neutrino mixing.

\subsubsection{Facilities for Three-Flavor Oscillations}

Facilities for measurements of three-flavor oscillation parameters include JUNO, Fermilab/SURF~\cite{Eldred:2022vxi,Ainsworth:2021ahm,Heise:2022iaf}, J-PARC/\-Kamioka~\cite{Endo:2022imj}, IceCube/DeepCore, and KM3NeT, with technically diverse characteristics. Current-generation experiments each have some sensitivity to the three primary oscillation physics goals, but are unlikely to reach the required statistical thresholds for definitive determinations. Next-generation experiments are expected to provide definitive answers to these questions for most of the three-flavor parameter space. The broad experimental program reflects the fact that there are many parameters in the three-flavor oscillation picture with partially degenerate effects on oscillation observables, such that multiple experiments measuring the same parameter via different oscillation channels and experimental techniques are highly desirable to disentangle three-flavor effects and check the completeness of the three-flavor paradigm.

\begin{figure}[thbp]
    \centering
    \includegraphics[width=0.5\linewidth]{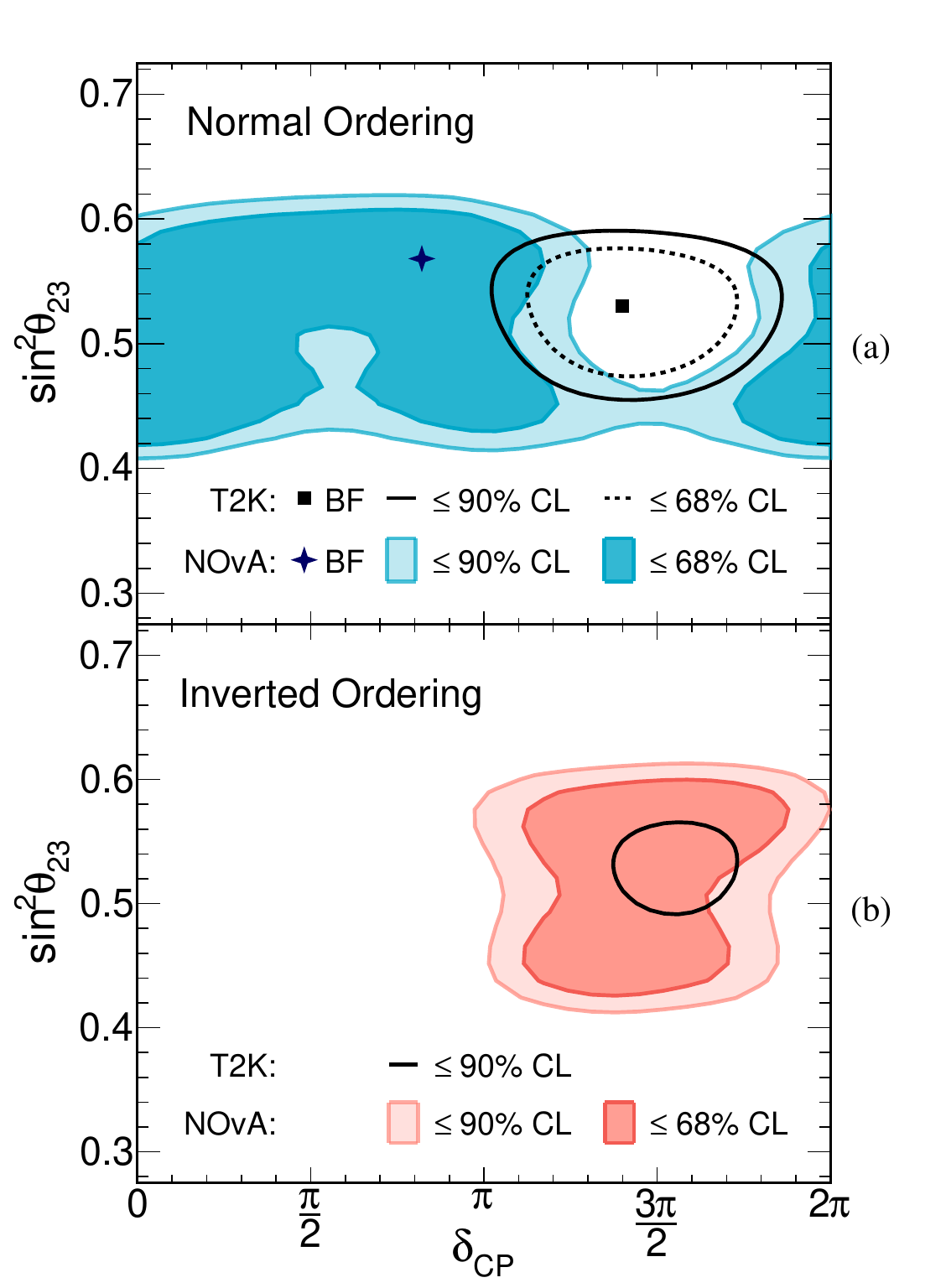}
    \caption{The best-fit values of $\sinst{23}$ and $\deltacp$ to NOvA and T2K
    data for the normal neutrino mass ordering (top), and the inverted 
    neutrino mass ordering (bottom)~\cite{NOvA:2021nfi}.}
    \label{fig:novat2kresults}
\end{figure}

There are currently two operating long-baseline experiments: NOvA in Minnesota with a beam from Fermilab~\cite{NOvA:2021nfi}, and T2K at Kamioka with a beam from J-PARC. Both measure \numu disappearance and \nue appearance in a narrow-band neutrino beam. NOvA measures neutrinos from the NuMI beam using functionally identical segmented liquid scintillator detectors located onsite at Fermilab and at a baseline of 810~km, in northern Minnesota. T2K measures neutrinos from the J-PARC beam using a suite of near detectors and the Super-Kamiokande water Cherenkov detector as the far detector at a baseline of 295 km. T2K and NOvA have been taking data since 2009 and 2014, respectively, have observed \nue and \anue appearance and have published measurements of \dm{32}, \sinstt{23}, and \deltacp. Figure~\ref{fig:novat2kresults} shows a comparison of the best-fit values for \sinstt{23} and \deltacp for the two experiments, using external data to constrain \sinstt{13}. T2K sees a relatively large asymmetry in \numutonue vs. $\bar{\nu}_\mu \rightarrow \bar{\nu}_e$, thus favoring CP-violating values of \deltacp, while NOvA does not see such an asymmetry, resulting in a slight tension between the experiments~\cite{NOvA:2021nfi}. A joint fit for oscillations is being undertaken by the two collaborations. 

Both experiments plan to continue taking data into the mid 2020s. For the most favorable parameters, NOvA expects up to 4$\sigma$ sensitivity to the neutrino mass ordering and T2K expects in the most optimistic case $>3 \sigma$ sensitivity to CP violation.  The actual sensitivity for both experiments is, however, highly dependent on the true values of the oscillation parameters, and it is not expected that this current generation of experiments will be able to provide definitive answers to the primary questions in the Neutrino Frontier.

Both Fermilab and J-PARC/Kamioka plan to host next-generation neutrino experiments with the goal of definitively addressing the three open questions in neutrino oscillation. DUNE~\cite{DUNE:2022aul} is the next step in the evolution of the Fermilab program. The DUNE collaboration was formed to realize the 2014 P5 vision of a best-in-class long-baseline experiment based at Fermilab. As the next-generation underground water Cherenkov detector in Japan, Hyper-Kamiokande (HK)~\cite{Hyper-Kamiokande:2022smq} builds on the highly successful Super-Kamiokande and T2K experiments.

\begin{figure}
    \centering
    \includegraphics[width=0.45\linewidth]{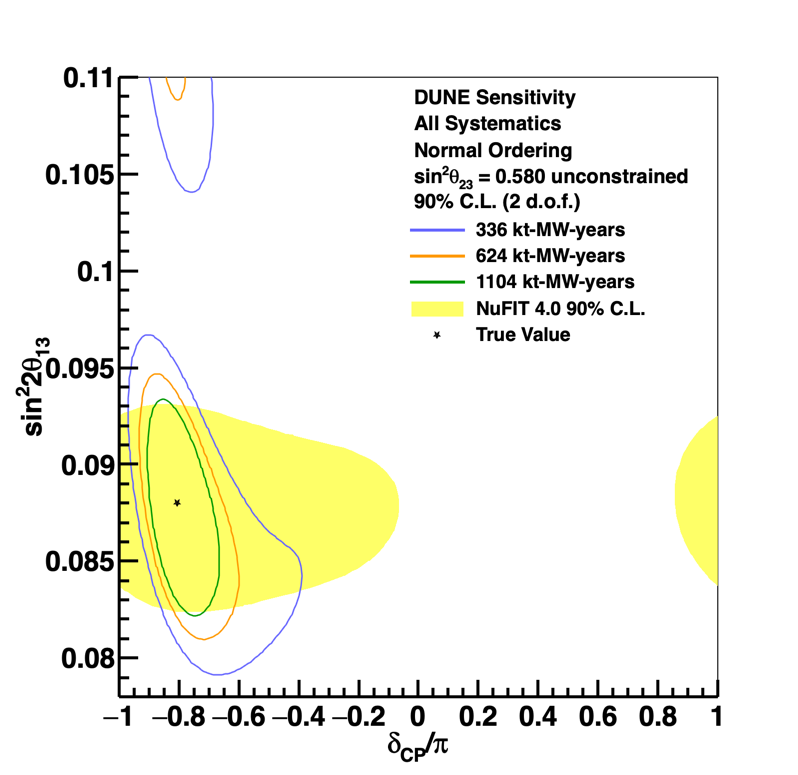}
    \includegraphics[width=0.45\linewidth]{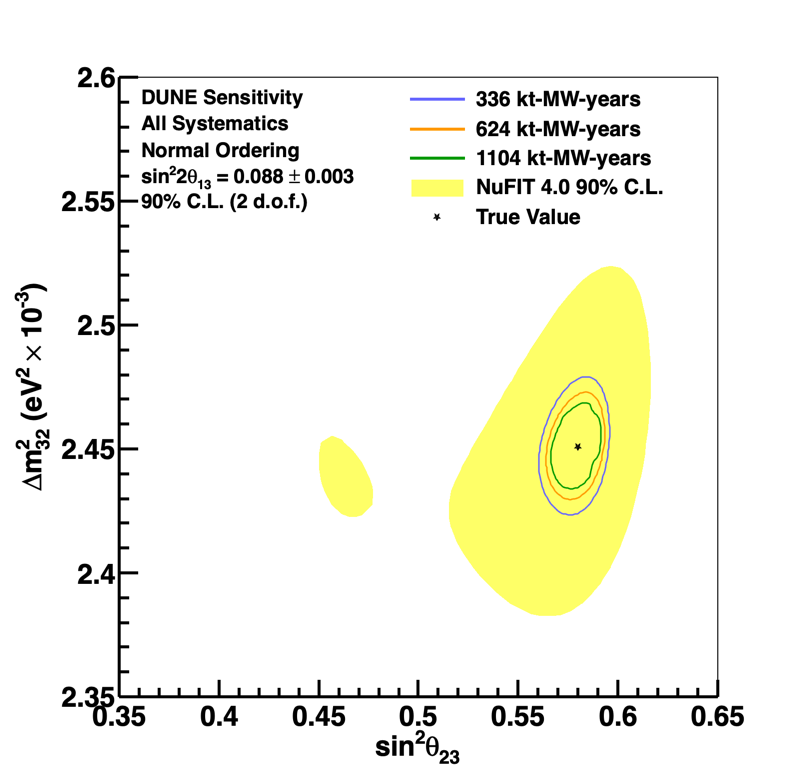}
    \caption{90\% confidence intervals for $\sinstt{13} - \deltacp$ (left), and $\sin^2{23} - \dm{32}$ (right) after a range of exposures in kt-MW-years, for a projected measurement with assumed true parameter values near the current global best fit. Yellow regions indicate recent global fits from NuFIT 4.0.}
    \label{fig:dunesensitivity}
\end{figure}

DUNE is designed to achieve its physics goals by pushing both the intensity and precision beyond what is achieved by current long-baseline neutrino oscillation experiments. Key features of this design include a high-power beam, a broad neutrino spectrum to measure the shape of the oscillation pattern, a long baseline of $\sim$1300~km, large far detectors located deep underground using liquid argon technology for particle identification and energy measurement, and an optimized near detector~\cite{DUNE:2021tad} with the same technology as the far detector to constrain systematics and enable a broad physics program. Critical components of the near detector are movable, to facilitate collection of data at multiple off-axis angles, which will reduce sensitivity to neutrino interaction modeling, a critical source of systematic uncertainty for long-baseline analyses; this capability is referred to as DUNE-PRISM~\cite{DUNE:2021tad}. DUNE’s long baseline provides excellent sensitivity to the matter effect on neutrino oscillations and hence strong resolving power for the neutrino mass ordering.  DUNE will be able to make precise measurements of \dm{32}, \sinst{23}, \sinstt{13}, and \deltacp. Figure~\ref{fig:dunesensitivity} shows the ultimate DUNE sensitivity to these parameters after several exposures. It is interesting to note that DUNE sensitivity to \sinstt{13} will ultimately be comparable to that from reactor experiments, allowing a unitarity test of the PMNS matrix.

As a result of resource limitations on the LBNF and DUNE projects, DUNE will be built in two phases, detailed in Table~\ref{tab:dunephases}. The highest priority is to deliver the full LBNF facility, including far site conventional facilities to accommodate four 17-kt FD (each at least 10-kt fiducial) modules, near site conventional facilities to support the full ND, and a 1.2 MW beam that can be upgraded to 2.4 MW. LBNF will be completed before the beginning of DUNE Phase I data-taking in the late 2020s. The Phase I configuration, consisting of the first half of the required far detector mass, half of the required beam power, and the minimal suite of ND components required to make credible oscillation physics measurements, is sufficient for early physics goals, including the determination of the neutrino mass ordering. However, Phase II is required to achieve the precision neutrino oscillation physics goals laid out by the last P5. Phase II, which completes the full DUNE scope, and includes upgrades to at least 40-kt of far detector fiducial mass, a 2.4 MW proton beam, and the full near detector, is critical for both the oscillation and non-oscillation physics programs of DUNE. Additionally, each of the Phase II upgrade components also provides an opportunity to broaden DUNE’s physics portfolio to address additional science drivers and to therefore involve a wider community, including potential opportunities for dark matter detection, neutrinoless double-beta decay measurements~\cite{Mastbaum:2022rhw}, and additional BSM searches. There is also significant community interest in other experimental efforts that may make use of Fermilab’s upgraded proton accelerator complex~\cite{Arrington:2022pon}.

\begin{table}[h]
    \centering
    \begin{tabular}{c|c|c|c}
        Parameter  & Phase I      & Phase II & Impact \\ \hline
        FD mass    & 20 kt fiducial       & 40 kt fiducial    & FD statistics \\
        Beam power & up to 1.2 MW & 2.4 MW   & FD statistics \\
        ND config  & ND-LAr,TMS, SAND          & TMS $\rightarrow$ ND-GAr~\cite{DUNE:2022yni} & Syst. constraints \\
    \end{tabular}
    \caption{A description of the two-phased approach to DUNE. ND-LAr, including the PRISM movement capability, and SAND are present in both phases of the ND.  TMS: Temporary Muon Spectrometer. SAND: System for on-Axis Neutrino Detection.}
    \label{tab:dunephases}
\end{table}

\begin{figure}
    \centering
    \includegraphics[width=0.9\linewidth]{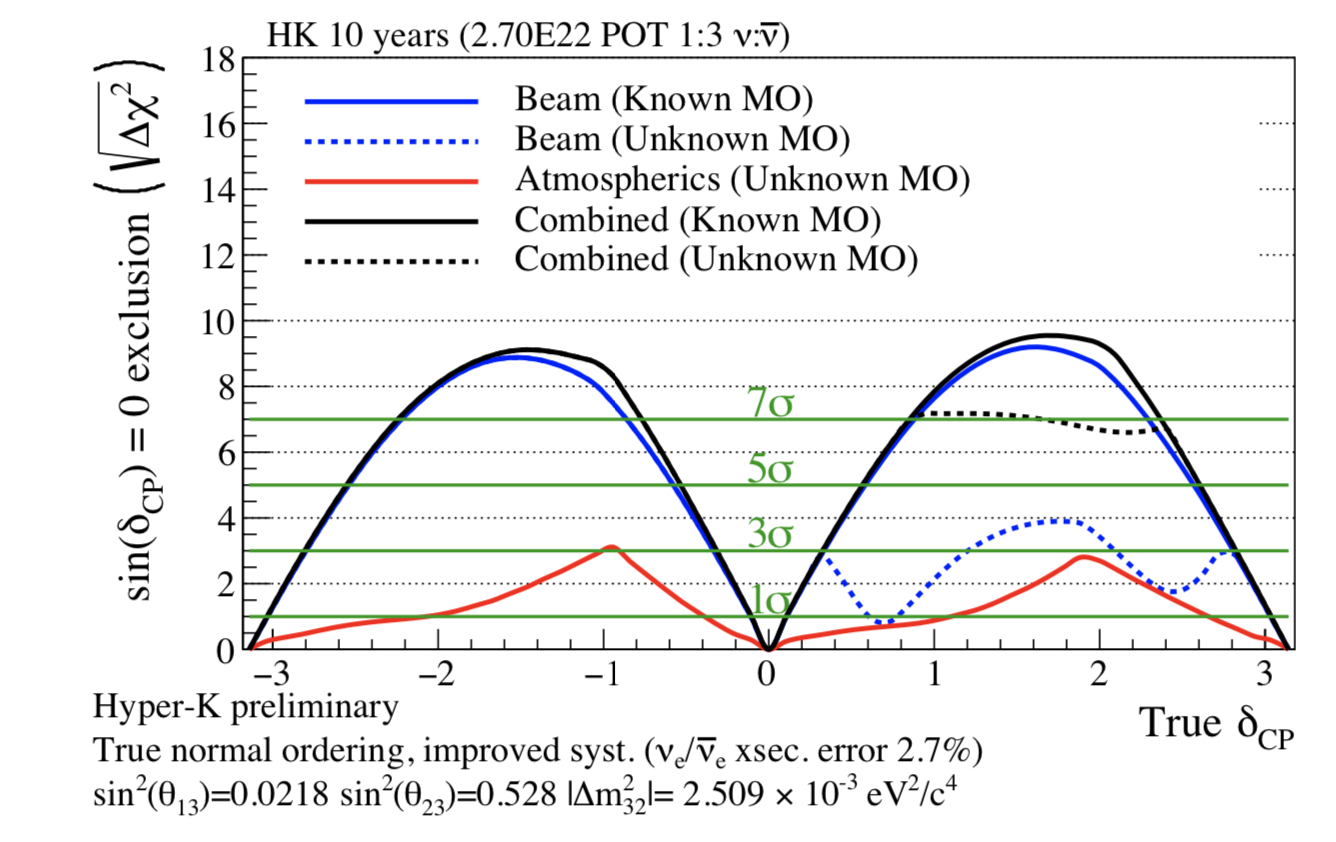}
    \caption{HK sensitivity to exclude $\sin\delta (cp) = 0$, plotted as a function of the true value of \deltacp, assuming the mass ordering is unknown. A combined fit of HK beam and atmospheric neutrinos significantly enhances the HK sensitivity to \deltacp.}
    \label{fig:hksens}
\end{figure}

The HK experiment plans to utilize the upgraded J-PARC 1.3~MW proton beam, an upgraded near detector, a proposed new intermediate detector to be installed $\sim$1 km away from the neutrino production target, and a new 260~kt far detector located 295 km away from the neutrino source near Super-K. The new intermediate detector would allow for measurements at a range of off-axis angles for use in a PRISM analysis. 
As shown in Figure~\ref{fig:hksens}, HK will have strong sensitivity to CP violation for much of parameter space, assuming the neutrino mass ordering is known. While HK's design parameters will make mass ordering determination difficult using its accelerator neutrino sample alone, a simultaneous fit to accelerator and atmospheric neutrinos is expected to significantly enhance sensitivity to oscillation parameters. As in DUNE, excellent control of systematic uncertainties is critical, with significant improvements in sensitivity possible if systematics are reduced relative to current T2K levels. In addition to the accelerator and atmopheric neutrino samples discussed here, HK will also be sensitive to a broad array of physics processes, including nucleon decay, solar neutrinos, supernova burst neutrinos, and diffuse supernova background neutrinos, with significant sensitivity to physics beyond the standard model.
Civil construction for the HK tank is underway, with data taking scheduled to start in 2027.

JUNO (Jiangmen Underground Neutrino Observatory) is a large liquid scintillator detector in South China. JUNO projects measurements of \sinstt{12}, \dm{21}, and \dm{32} with resolutions $\sim \pm 1$\% in six years of data taking, using reactor antineutrinos from two commercial power plants, detected via inverse beta decay. If JUNO can achieve the goal energy resolution of $\pm$3\%, the question of
the neutrino mass ordering can be resolved at 3$\sigma$ by observing the phase of the “fast” oscillation from \dm{32} superimposed on the “slow” oscillations due to \dm{21} in the recoil electron energy spectrum. JUNO anticipates that data collection will begin in 2023.

IceCube can observe atmospheric neutrinos in a very wide energy range, from the order of GeV to 100 TeV. The IceCube Upgrade improves the resolution of neutrino energy and direction and lowers the energy threshold, allowing for improved neutrino oscillation sensitivity in atmospheric neutrinos. In particular, the measurement of the matter effect of atmospheric neutrinos by IceCube Upgrade, combined with the JUNO experiment, has significant sensitivity to the neutrino mass ordering. Atmospheric neutrinos, particularly in the high-energy region, are also well suited for searching for physics beyond the standard model.

\subsubsection{Supporting Measurements and Capabilities}

Auxiliary measurements, particularly hadron production measurements to constrain neutrino fluxes and flux systematic errors, as well as precision measurements of cross-sections to constrain interaction-model systematic errors, are necessary for precision three-flavor oscillation measurements. 
Some of these essential measurements are described in Sections \ref{sect:interactions} and \ref{sect:hornbeams} in this document and in more detail in reports from NF06 (Neutrino Interaction Cross Sections)\cite{Balantekin:2022jrq} and NF09 (Artificial Neutrino Sources)\cite{Fields:2022pxk}.

\subsection{Physics Beyond the Standard Model in Neutrinos and Neutrino Experiments}

The arrival of the ``precision era'' in neutrino physics comes with both challenges and opportunities.  One of these opportunities is the ability to search for evidence of or to discover BSM physics, expanding the physics scope of current and future experiments well beyond the measurement of three-flavor neutrino oscillations. In fact, while the SM describes with great accuracy most of the phenomena observed in high-energy physics, a number of fundamental scientific questions still need to be answered:
\begin{enumerate}
    \item What are the properties and interactions of neutrinos, and how many neutrinos are there? What is the explanation behind the observed anomalies in short-baseline neutrino experiments? 
    \item What is the mechanism responsible of neutrino mass generation? Are lepton and baryon numbers fundamentally conserved, or just accidental symmetries? 
    \item What is the mechanism responsible for the matter-antimatter asymmetry of the Universe?
    \item Is the dark sector as complex and phenomenologically rich as the visible sector? How does the dark sector interact with the visible sector? 
    \item Do all the forces unify? If so, at what energy scale? 
\end{enumerate}

Current and future neutrino experiments are in an excellent position to explore many of the open problems in the SM, and significant progress has been made in addressing the questions outlined above in the past decade. While the Snowmass 2013 neutrino working group report~\cite{IntensityFrontierNeutrinoWorkingGroup:2013sdv} did include a section on BSM opportunities, it was largely focused on resolving the short-baseline anomalies through the addition of an eV-scale light sterile neutrino~\cite{Acero:2022wqg}. Since then, the community has made a significant effort in pushing the boundaries and the potential of neutrino experiments to test for BSM models more broadly, and expanding searches to non-minimal BSM scenarios in the neutrino sector.

\subsubsection{Heavy Neutral Leptons, Sterile Neutrinos, and the Short-Baseline Anomalies}

Heavy Neutral Leptons (HNLs) are right-handed neutrino partners to the standard model active neutrinos. Their existence can provide elegant solutions to present open questions in fundamental physics such as the origin of neutrino masses, the nature of dark matter and the observed matter antimatter asymmetry in the Universe. In the minimal SM extensions, these HNLs, named as such because they are significantly heavier than the standard model active neutrinos, are (quasi) sterile and are produced through mixing with the active neutrinos; as such, they are also commonly referred to as sterile neutrinos, or simply as right-handed neutrinos. However, it is important to bear in mind that, in non-minimal models, they may possess additional interactions (for example, in extensions of the SM with additional gauge symmetries such as B-L, or in non-minimal dark sector models). Depending on their masses, the existence of such heavy neutrinos can lead to multiple phenomenological consequences in a variety of experiments, from neutrino oscillation measurements to colliders (e.g.,\cite{Batell:2022ubw,Mekala:2022cmm}). In this section we first briefly review the impact of sterile neutrinos on short-baseline neutrino oscillation searches and the reported neutrino anomalies. We will then discuss the phenomenological implications due to HNLs at heavier scales.

The past three decades of experimental neutrino measurements have accumulated hints of observed anomalous short-baseline flavor transformation from multiple sectors of varied neutrino source (proton accelerators, reactors, and intense radioactive sources) and energy (from MeV to GeV scales). These are commonly referred to as the LSND Anomaly, the MiniBooNE Low-Energy Excess, the Reactor Antineutrino Anomaly, and the Gallium Anomaly. Some of these observations have grown in significance over time to nearly or over $5\sigma$, and all share the commonality of having been observed at very short neutrino propagation distances, $L\sim \mathcal{O}(1\,\mathrm{m} - 1\,\mathrm{km})$, relative to neutrino production, i.e.,~before any three-neutrino oscillation effects are expected to become measurable. These anomalies thus serve as an intensifying experimental impetus for pursuing BSM physics accessible through the neutrino sector.  With the generation mechanism of neutrino mass currently unknown, the search for new neutrino mass states and hidden-sector couplings is also theoretically well-motivated.

\renewcommand{\arraystretch}{1}
\newcommand{\cmark}{\ding{51}}%
\newcommand{\xmark}{\ding{55}}%
\newcommand{\nocheck}{\textcolor{red}{\xmark}}
\newcommand{\semicheck}{\textcolor{orange}{\cmark}}
\newcommand{\fullcheck}{\textcolor{green}{\cmark}}
\newcommand{\newhline}{\cline{2-8}}

\begin{table*}[!htp]
    \centering
    \begin{sideways}
\resizebox{20cm}{!}{%
    \begin{tabular}{|>{\centering\arraybackslash}p{3 cm}|>{\centering\arraybackslash}p{4 cm}|>{\centering\arraybackslash}p{3.2 cm}|>{\centering\arraybackslash}p{2 cm}|>{\centering\arraybackslash}p{1.9 cm}|>{\centering\arraybackslash}p{1.4 cm}|>{\centering\arraybackslash}p{1.3 cm}|>{\centering\arraybackslash}p{2.5 cm}|}
\hline\hline
    \multirow{2}{*}{Category} & \multirow{2}{*}{Model}& \multirow{2}{*}{Signature} & \multicolumn{4}{c|}{Anomalies} & \multirow{2}{*}{References} \\ \cline{4-7} & & & LSND & MiniBooNE & Reactor & Gallium & \\
\hline\hline
\multirow[c]{3}{*}[-3em]{ \parbox[t]{3 cm}{\vspace{-0.9cm}\centering \textbf{Flavor Conversion:} Transitions }}
        & (3+N) oscillations & oscillations & \fullcheck & \fullcheck & \fullcheck & \fullcheck & Reviews and global fits \cite{Dentler:2018sju,Diaz:2019fwt,Boser:2019rta,Dasgupta:2021ies}
        \\
        \newhline
        & (3+N) w/ invisible sterile decay & oscillations w/ $\nu_4$ invisible decay &  \fullcheck &  \fullcheck & \fullcheck & \fullcheck & \cite{Moss:2017pur,Moulai:2019gpi}
        \\
        \newhline
        & (3+N) w/ sterile decay & $\nu_4 \to \phi \nu_e$ &  \fullcheck &  \fullcheck & \semicheck & \semicheck & \cite{Palomares-Ruiz:2005zbh,Bai:2015ztj,deGouvea:2019qre,Dentler:2019dhz,Hostert:2020oui}
        \\
\hline\hline
\multirow[c]{2}{*}[-2em]{ \parbox[t]{3 cm}{\vspace{-0.5cm}\centering \textbf{Flavor Conversion:} Matter Effects }}
        & (3+N) w/ anomalous matter effects  & $\nu_\mu \to \nu_e$ via matter effects & \fullcheck  & \fullcheck & \nocheck & \nocheck & \cite{Akhmedov:2011zza,Bramante:2011uu,Karagiorgi:2012kw,Asaadi:2017bhx,Smirnov:2021zgn}
        \\ 
        \newhline
        & (3+N) w/ quasi-sterile neutrinos  & $\nu_\mu \to \nu_e$ w/ resonant $\nu_s$ matter effects & \fullcheck & \fullcheck & \semicheck & \semicheck & \cite{Alves:2022vgn}
        \\
\hline\hline
\multirow[c]{2}{*}[-2em]{ \parbox[t]{3 cm}{\vspace{-0.5cm}\centering \textbf{Flavor Conversion:} Flavor Violation  }} 
        & lepton-flavor-violating $\mu$ decays & $\mu^+\to e^+ \nu_{\alpha}\overline{\nu_e}$ & \fullcheck & \nocheck & \nocheck & \nocheck & \cite{Bergmann:1998ft,Babu:2016fdt,Jones:2019tow}
        \\ 
        \newhline
        & neutrino-flavor-changing bremsstrahlung & $\nu_\mu A \to e \phi A$ & \fullcheck & \fullcheck & \nocheck & \nocheck  & \cite{Berryman:2018ogk} 
        \\
\hline\hline
\multirow[c]{2}{*}[-2em]{ \parbox[t]{3 cm}{\vspace{-0.7cm}\centering \textbf{Dark Sector:} Decays in Flight }}
        & transition magnetic mom., heavy $\nu$ decay & $N\to \nu \gamma$ &  \nocheck & \fullcheck & \nocheck & \nocheck & \cite{Fischer:2019fbw}
        \\
        \newhline
        & dark sector heavy neutrino decay & $N\to \nu (X \to e^+e^-)$ or $N\to \nu (X\to \gamma \gamma)$ & \nocheck & \fullcheck & \nocheck & \nocheck & \cite{Chang:2021myh}
        \\
        \newhline
\hline\hline
\multirow[c]{2}{*}[-2em]{\parbox[t]{3 cm}{\vspace{-0.7cm}\centering \textbf{Dark Sector:} Neutrino Scattering }} 
    & neutrino-induced up-scattering & $\nu A \to N A$,  $N\to \nu e^+e^-$ or $N \to \nu \gamma \gamma$ & \semicheck & \fullcheck & \nocheck  & \nocheck & \cite{Bertuzzo:2018itn,Bertuzzo:2018ftf,Ballett:2018ynz,Ballett:2019pyw,Datta:2020auq,Dutta:2020scq,Abdullahi:2020nyr,Abdallah:2020biq,Abdallah:2020vgg,Hammad:2021mpl}
    \\
    \newhline
    & neutrino dipole up-scattering & $\nu A \to N A$,  $N\to \nu \gamma $ & \semicheck & \fullcheck & \nocheck & \nocheck  & \cite{Gninenko:2009ks,Gninenko:2010pr,Gninenko:2012rw,Masip:2012ke,Radionov:2013mca,Magill:2018jla,Vergani:2021tgc,Alvarez-Ruso:2021dna}
    \\
\hline\hline
\multirow[c]{2}{*}[-2em]{ \parbox[t]{3 cm}{\vspace{-0.9cm}\centering \textbf{Dark Sector:} Dark Matter Scattering }}
    & dark particle-induced up-scattering  & $\gamma$ or $e^+e^-$ & \nocheck & \fullcheck & \nocheck & \nocheck & \cite{Dutta:2021cip}
    \\
    \newhline
    & dark particle-induced inverse Primakoff & $\gamma$ & \fullcheck & \fullcheck & \nocheck & \nocheck & \cite{Dutta:2021cip}
    \\
\hline\hline
\end{tabular}
}
 \end{sideways}
    \caption{New physics explanations of the short-baseline anomalies categorized by their signature. Notation: \fullcheck -- the model can naturally explain the anomaly,  \semicheck -- the model can partially explain the anomaly,   \nocheck -- the model cannot explain the anomaly.~\label{tab:sec-3:big_picture}}
\end{table*}
%%%%%%%%%%%%%%%%%%%%%%%%%%%%%%%%%%%%%%%%%%%%%%%%%%%%%%%%%%%%%%%%%%%%%

Over the past ten years, the neutrino community has made significant progress in implementing the primary recommendations of the previous P5 Report by developing a diverse program of small-scale experiments aimed at directly addressing these short-baseline anomalies and at probing the leading (at the time) theoretical explanation for their existence: oscillations from a single eV-scale sterile neutrino state, or a `3+1' scenario.  

The community's accomplishments have also extended to confronting the 3+1 oscillation explanation with complementary datasets from MicroBooNE, MINOS/MINOS+, IceCube, T2K, NOvA, Super-K, KATRIN, PROSPECT, and others. Importantly, motivated by the growing tension of experimental data sets under a 3+N light sterile neutrino oscillation hypothesis prompting the consideration of alternate viable anomaly interpretations, the community has additionally developed a rich array of new physics models beyond 3+N sterile neutrino oscillation scenarios. Those models range from exotic flavor transformation ones to dark sector models, many involving HNLs more broadly, and with rich phenomenology accessible at current, planned, and future short-baseline and other experiments. A summary of the anomaly interpretations is provided in Table~\ref{tab:sec-3:big_picture}, mapping specific models onto broader model classes and onto which anomalies they are meant to address.  

While the underlying source of the anomalies remains unresolved, the improvement in knowledge over the past decade is significant, and compels the deeper probes that are possible with current and upcoming experimental capabilities and facilities.  
Whether the anomalies are caused by a unified, but complex, regime of new physics, BSM effects specific to subsets of neutrino source types, or systematics, the adoption of a multi-probe test strategy is essential to their convincing resolution.  
Thanks to joint efforts by the theoretical and experimental communities, the path forward to the resolution of the anomalies is clearer, as is the possible range of exciting new physics scenarios that can be searched for with current and future facilities.  

A key ingredient to unearthing the origins of the short-baseline neutrino anomalies during the next decade is continued support for and full exploitation of short-baseline experiment investments made over the past decade.  
To that end, measurements by operating or imminent experiments will provide direct experimental tests of most of the anomalies and new information during the first half of the coming P5 period, which will enable tests of many popular interpretations with high sensitivity.  
This program represents a large international commitment with several of the involved experimental efforts sited at U.S. national laboratories.

Additionally, sustained community effort over the past decade has been expended to develop new experimental concepts, including cutting-edge sources and/or techniques for probing the primary BSM-related anomaly origin categories -- flavor transformation phenomena and new particle production -- with significantly improved sensitivity over the aforementioned experimental efforts.  
In some cases, advanced levels of currently achieved technical maturity could enable timely installation and completion of new experimental concepts, potentially providing a new wave of valuable short-baseline information well before the close of the coming P5 period.  
Other concepts relying on facilities or detectors that may be available later in the coming P5 period can sustain this new wave well into the following decade.

As examples, precise new BSM flavor transformation measurements can be made in the electron disappearance channel using precision measurements of beta decay and electron capture, isotope decay-at-rest sources, or large-volume short-baseline reactor or radioactive source experiments; in the muon disappearance channel using atmospheric neutrinos or neutrinos from decays of accelerator-produced kaons; and using well-characterized flavor content 
at future long-baseline oscillation facilities or with CEvNS-based measurements of decay-at-rest neutrinos.  
Furthermore, a vast array of potential dark sector couplings can be probed with high-statistics datasets from intense decay-in-flight, decay-at-rest, and collider beam dump neutrino sources or with atmospheric neutrino fluxes.  

Meanwhile, it will be important to continue to investigate complementary and/or indirect probes offered through currently-operating long-baseline accelerator measurements, atmospheric and solar neutrino constraints, precision measurements of weak nuclear decay products, neutrinoless double beta decay measurements, and constraints from cosmology.  The recently confirmed Gallium Anomaly, a more than 5\,$\sigma$ deficit of electron neutrinos, poses a major riddle worth pursuing.

Full exploitation of the above investments provides plausible paths to convergence of interpretation of the LSND, MiniBooNE, and Reactor Anomalies optimistically as early as the first half of the coming P5 period.  
Should the interpretations converge, the short-baseline experimental and theoretical landscape will shift, with subsequent priorities depending heavily on what is learned about the various hypothesized anomaly origins.  
Regardless of what is discovered,
we should be prepared to re-evaluate priorities given what will be learned. 
It is therefore essential that available supporting resources are structured to enable adaptation to developments during the next P5 period and to ensure that timely experimental groundwork is laid for following generations of precision short-baseline measurements.

Besides neutrino oscillation measurements, the existence of HNL can be probed in multiple ways, including: reactor experiments searching for HNL decays in flight, or tests of energy-momentum conservation in nuclear reactions (sensitive to masses between $\sim \mathrm{keV}$ and 10 MeV); fixed-target facilities (giving the best sensitivities to HNLs in the tens of MeV to few GeV range in the near future), a broad category that can be further divided into searches from rare kaon decays, beam-dump setups, and searches in accelerator-neutrino-beam environments; and collider experiments (sensitive to HNLs with masses above a few GeV). In the latter case it is worth mentioning the recent development of a series of proposals for transverse (MATHUSLA, CODEX-b, AL3X, ANIBUS and MAPP-LLP) and forward (FASER, SND@LHC, the Forward Physics Facility~\cite{Anchordoqui:2021ghd} and FACET) facilities at the LHC, which provide enhanced sensitivities to HNL in the GeV to tens of GeV mass region.  In addition to these terrestrial bounds, a landscape of constraints arises from solar, atmospheric, astrophysical, and cosmological considerations. At present, solar neutrino up-scattering searches provide strong constraints on minimal HNLs, while atmospheric neutrino up-scattering provides novel sensitivity to non-minimal HNLs (such as those interacting with a transition magnetic moment, for example). Likewise, the presence of HNLs in the early universe can be strongly constrained given that they can disrupt the success of big bang nucleosynthesis. A summary of the main phenomenological signals in HNL models is provided in Tab.~\ref{tab:big_table_prospects}.

It is worth stressing that for most of the future options and proposals in the literature, only first estimates on the sensitivity to HNL have been made. Although these demonstrate the potential HNL parameter space coverage, certainly further studies are strongly desirable. Likewise, it would be desirable for future collider facilities to take into account from the start the strong interest and need for searches for long-lived particles in their infrastructure plans, so that detector designs may include searches for HNLs in their baseline physics targets from the start. 

As a final remark, it is worth noting the significantly different sensitivities and constraints for non-minimal HNLs scenarios.  The phenomenological implications may change significantly once the single-flavor dominance hypothesis is abandoned, or if the HNL possesses new interactions (either with the visible or the dark sector).

\subsubsection{BSM Signals in the Neutrino Sector: Neutrino Flavor and Neutrino Scattering}

Neutrino oscillation experiments are very precise interferometers, sensitive to subleading effects from new physics affecting neutrino flavor transitions. On a separate front, neutrino telescopes -- detectors of astrophysical neutrinos -- provide a unique avenue to probe BSM effects, given the very long distances traveled by the detected neutrinos (which range from the Earth radius to several gigaparsecs), as well as their ultra-high energies. Finally, astrophysical observations (such as nearby core-collapse supernovae) in the upcoming decade may provide invaluable information and allow us to test neutrino propagation in extremely dense environments. A summary of the main phenomenological scenarios that can lead to observable signatures in neutrino oscillations and neutrino flavor transitions, together with the most relevant experimental signatures for each of them, is provided in Tab.~\ref{tab:big_table_prospects}. For reference, the table also includes a list of some of the planned/proposed experiments where such signals may be investigated (see the provided references for further details). 

In the future, a precise measurement of neutrino oscillations in different environments (in matter and vacuum,  on different oscillations channels, and/or for experiments relying on different detection mechanisms) will be required to ensure that the three-neutrino paradigm is robust. 
Because of this, global fits between different data sets may be critical to unveil new physics effects on neutrino flavor, in the same way that they provided the first evidence for a non-zero $\theta_{13}$ before it was experimentally measured. The same argument also extends to non-oscillation measurements as well: as a well-known example, CEvNS and oscillation data are strongly complementary in setting bounds on new neutrino interactions, but the full potential can only be unlocked from a joint fit to all available data.

Future advances in this direction will require a joint effort between the experimental and theoretical communities, though. On one hand, global fits rely on experimental collaborations facilitating as much information as possible. IceCube and COHERENT are two notable examples where data release efforts have empowered the theoretical community to derive new bounds on a variety of BSM scenarios, and/or to propose new experimental searches for new BSM signals; however, for many other experiments there is very little information publicly available. On the other hand, the theoretical community often makes use of theoretical frameworks and parameterizations to derive bounds on new physics (a common example is the use of effective operators at low energies, an approach commonly used in other areas of particle physics as well). However, it is equally important that these frameworks  eventually be matched onto viable models, so that the bounds from neutrino experiments can be contrasted with those obtained in other Frontiers of particle physics. 

At the same time, while the potential of neutrino facilities for BSM searches is clear, it should be kept in mind that in the past many of these have traditionally been impractical (if not impossible), mainly due to the large systematic uncertainties stemming from our imprecise knowledge of both neutrino fluxes and neutrino interaction cross sections. If we want to fully exploit future experimental capabilities to search for BSM signals, dedicated efforts are required to reduce systematic uncertainties according to the specific needs in each case.
The program of supporting measurements in the Neutrino Frontier, including hadron-production measurements for neutrino flux predictions and measurements to improve modeling of neutrino-nucleus interactions, is as important for BSM searches as it is for the standard three-flavor neutrino program. Similar considerations apply to neutrino telescopes: in this case, improving neutrino flavor identification will be key in order to improve experimental capabilities to test for BSM effects, and this will demand a dedicated effort.

Before moving on to the next section let us stress that, while non-standard interactions with charged SM fermions are in some cases straightforwardly tested with traditional $\nu$ scattering and oscillation experiments, neutrino self -interactions, $\nu$SI, can also be tested by a variety of methods. In particular, cosmological observables (including light element
abundances, the Cosmic Microwave Background, and the matter distribution in the Universe) can
probe self-interactions at scales of $\sim$eV to $\sim$MeV~\cite{Berryman:2022hds}. Supernovae and other astrophysical neutrino sources can test characteristic self-interaction scales up to $\mathcal{O}(100)$~MeV. Finally, laboratory experiments can access the broadest range of self-interaction scales all the way up to $\mathcal{O}(100)$~GeV.

\subsubsection{Neutrino Experiments and Signals from BSM Physics in Other Sectors: Dark Matter and Baryon Number Violation}

Besides the origin of neutrino masses and mixing, neutrino experiments may also be able to shed some light on two of the other major open problems in the SM: the nature of dark matter, and the origin of the baryon asymmetry of the Universe. 

The idea of a dark sector comprising new states that are not charged under the known forces, but are weakly coupled to the standard model via a new mediator, is well motivated from a variety of perspectives. In fact, arguably the dominant empirical motivations for BSM physics searches concern neutrino mass and dark matter: their phenomenology points primarily to weak coupling, but not to a specific new physics scale. Nevertheless, the concept of dark matter direct detection was strongly motivated by the WIMP scenario. In this regard it is worth noting that a wide range of new, well-motivated physics models and dark-sector scenarios have been proposed in the last decade, which would explain the absence of a signal in direct detection experiments, and at the same time may lead to a new physics signals elsewhere. 

Modern accelerator-based neutrino beam facilities feature enormous proton beam-target collision luminosities, which can supply copious secondary forward fluxes of dark sector particles. Thus, neutrino beam experiments allow for BSM physics searches in a way that is complementary to Energy Frontier experiments. At the same time, well-motivated physics models for non-minimal dark sectors predict cosmogenic signals complementary to those in conventional direct DM detection experiments. These often have less intense and more energetic fluxes, to which underground, kiloton-scale neutrino detectors can be readily sensitive. Consequently, both the theoretical and experimental neutrino communities are now heavily exploring the potential of neutrino facilities to test for signals from non-minimal dark sectors. The landscape of potential signals at neutrino experiments coming from BSM physics in other sectors is wide, as summarized in Table~\ref{tab:big_table_prospects}.

On a separate front, neutrino experiments have been (and will be) widely used to search for Baryon Number Violation (BNV). Although there is no direct evidence of proton decay so far, the observed matter-antimatter asymmetry suggests that baryon number must be violated at some level. While the main focus of BNV experiments is on proton decay searches, other equally important baryon and/or lepton number violating processes include dinucleon decays and neutron-antineutron oscillations, which must be studied as well.
Regarding experimental detection prospects,  detector mass is obviously a clear and crucial characteristic in next-generation BNV searches, and clearly Hyper-Kamiokande has the advantage in that respect. However, detector technology will also play an important role. In fact, DUNE's excellent imaging capabilities and JUNO's superb 
resolution offer advantages in some channels over Hyper-Kamiokande's larger mass. The proposed Theia~\cite{Theia:2022uyh} experiment would combine the advantages of the large mass of a water Cherenkov detector with the good resolution of a liquid scintillator detector. With this worldwide program, should a BNV signal be observed by any one detector in the next generation, confirmation from other detectors using different technologies would provide powerful evidence of new physics~\cite{Dev:2022jbf}.

In order to make significant progress in any of the directions outlined above in the next decade, BSM searches should be considered a high priority in the experimental design decisions for the future. A closer collaboration with the nuclear physics community would also be needed, as well as steps to enhance simulation tools for BSM processes and related backgrounds. Finally, new physics searches at neutrino facilities would significantly benefit from a closer collaboration across particle physics frontiers, given the strong complementarity with respect to similar efforts in other areas. It would also be desirable that experimental sensitivities and constraints be presented in a model-independent way as much as possible, to ease reinterpretation and recasting of the obtained results to other frameworks or BSM scenarios.

%\afterpage{%
    \clearpage% Flush earlier floats (otherwise order might not be correct)
    \thispagestyle{empty}% empty page style (?)
   
    \begin{table}
   
     \begin{sideways}
    %\captionsetup{font=ninept}
     \resizebox{1\textwidth}{!}{
     % \fontsize{10}{8}\selectfont
     \begin{tabular}{|l|l|l|l|}
    \hline
    BSM Scenario &  Sources & Signatures  & Example Experiments 
    \\
    \hline 
    \hline
    \multirow{5}{*}{HNL\cite{Abdullahi:2022jlv}} 
    & Colliders  & HNL decay &  ATLAS, CMS, FASER, Belle II, ...  \cr  
    & Nuclear decays  & Nuclear decay kinematics & KATRIN/TRISTAN, HUNTER ...  \cr 
    & Fixed target & HNL decay & DUNE ND, SHiP, ICARUS, ... \cr \cline{2-4} 
    & \multirow{2}{*}{Atm. \& solar $\nu$s} & Distorted recoil spectrum & \multirow{2}{*}{DUNE, HK, IceCube/DeepCore, ...} \cr  
    & & HNL decay, double bangs & \cr \cline{2-4}
    & Early Universe  & Cosmological parameters ($N_{\rm eff}$) & Simons Observatory, CMB-S4, ... 
    \\
    \hline
    Non-unitarity\cite{Arguelles:2022xxa} &  Beam \& Atm. $\nu$s  & Deviations from 3-$\nu$ mixing (ND \& FD) & DUNE, ESS$\nu$SB, HK, ... \\
    \hline
    \multirow{3}{*}{LED\cite{Arguelles:2022xxa}} & 
    Reactor $\nu$s & \multirow{2}{*}{Distortion of oscillated spectra (FD \& ND)} & JUNO, TAO,...\cr 
    & Beam $\nu$s & & DUNE, ... \cr \cline{2-4}
    & Atm. $\nu$s & Anomalous matter effects & Icecube, KM3NeT, ... 
    \\
    \hline
     & 
    Reactor \& Stopped-$\pi$ sources & Distortion of CEvNS rate & COHERENT, CONNIE, CONUS, ... \cr 
    NSI \& light & Solar, Beam, Atm \& SN $\nu$s & Anomalous matter effects & DARWIN, DUNE, T2HKK, HK, IceCube, ...\cr 
     mediators\cite{Arguelles:2022xxa,Abdullah:2022zue} & Beam $\nu$s & Anomalous appearance, $\nu-e^-$ scattering, tridents & DUNE ND, T2HK ND, IsoDAR, ... \cr
    & Collider $\nu$s & Distortion of CC spectrum & FASER$\nu$, ... 
    \\
    \hline 
    Long-range & Solar \& Atm $\nu$ & Anomalous matter potential & HK, JUNO, DUNE, ... \cr 
    forces~\cite{Arguelles:2022xxa} & UHE Astrophysical nus & Distorted flavor ratios & HE Neutrino Telescopes 
    \\ 
    \hline
    \multirow{3}{*}{$\nu$-DM interact. \cite{Arguelles:2022xxa}} & Reactor \& solar $\nu$s & Distorted oscillated spectra, or & JUNO, ... \cr 
    & Beam $\nu$ & time-dependent oscillation params. & DUNE, ... \cr 
    & UHE Astrophysical $\nu$s & Distorted flavor ratios \& spectra & HE \& UHE Neutrino Telescopes
    \\
    \hline
     & SN $\nu$s & SN extra energy loss, distortion in neutrino spectra & DUNE, HK, JUNO, ... \cr
    $\nu$ self & UHE Astrophysical $\nu$s & {Distorted spectra} & HE \& UHE $\nu$ telescopes \cr
    interact.~\cite{Berryman:2022hds, Feng:2022inv} & Early Universe & Effects on CMB, BBN, \& structure formation & CORE, PICO, CMB-S4 \cr 
    & Beam \& Collider $\nu$s & Missing energy \& $p_T$ in $\nu$ scattering & DUNE ND, Forward Physics Facility, ...
    \\
    \hline
    \multirow{4}{*}{$\nu$ decay\cite{Arguelles:2022xxa}} & Reactor $\nu$s & \multirow{3}{*}{Distortion of oscillated spectra} & JUNO, ... \cr 
    & Beam $\nu$s  & & DUNE, MOMENT, ESS$\nu$SB, HK, IsoDAR, ...\cr 
    & Atm. $\nu$s  & & INO-ICAL, KM3NeT-ORCA, ... \cr \cline{2-4}
    & UHE Astrophysical $\nu$s & Distorted flavor ratios \& spectra & HE \& UHE Neutrino Telescopes 
    \\
    \hline
    \multirow{3}{*}{CPT violation\cite{Arguelles:2022xxa}} & Beam $\nu$s & \multirow{2}{*}{Different $\nu$ and $\bar{\nu}$ osc. params.} & DUNE, ESS$\nu$SB, HK, ... \cr 
    & Atm. $\nu$s & & IceCube, DUNE, ... \cr \cline{2-4}
    & UHE Astrophysical $\nu$s & Distorted flavor ratios \& spectra & HE \& UHE Neutrino Telescopes  
    \\
    \hline
    \multirow{3}{*}{Lorentz violation\cite{Arguelles:2022xxa}} & Beam $\nu$s & \multirow{2}{*}{Sidereal modulation of event rate} & DUNE, ESS$\nu$SB, HK, ... \cr 
    & Atm. $\nu$s & & IceCube, DUNE, ...  \cr \cline{2-4}
    & UHE Astrophysical $\nu$s & Distorted flavor ratios \& spectra, velocity dispersion & HE \& UHE Neutrino Telescopes 
    \\
    \hline
    \multirow{3}{*}{Quantum decoh.~\cite{Arguelles:2022xxa}} & Beam $\nu$s & \multirow{2}{*}{Distortion of oscillated spectra} & DUNE, IsoDAR, ... \cr 
    & Atm. $\nu$s & & KM3NeT, IceCube, HK, ...  \cr \cline{2-4}
    & UHE Astrophysical $\nu$s & Distorted flavor ratios & HE Neutrino Telescopes 
    \\
    \hline
    $B$ violation\cite{Dev:2022jbf} & Detector mass & Nucleon decay, $n-\bar{n}$ oscillations & DUNE, HK, JUNO, ...
    \\
    \hline
    \multirow{4}{*}{Dark Matter\cite{beambasedDM,cosmicDM}} 
    &  DM annihilation, DM decay & Excess of $\nu$s from Sun or Earth & \multirow{2}{*}{HK, DUNE, IceCube ...} \cr 
    & Boosted DM, slow-moving DM & Scattering, or up-scattering \& decay & \cr 
    \cline{2-4}
    & \multirow{2}{*}{Fixed target} &  Decay &  DUNE, T2HK, SBN, FASER$\nu$, ... \cr
    & & Scattering, or up-scattering \& decay & PIP2-BD, SBN-BD ...
    \\
    \hline 
    \multirow{2}{*}{Milli-charged particles\cite{beambasedDM}} & Fixed target & \multirow{2}{*}{Scattering} & DUNE ND, T2HK ND, IsoDAR, ... \cr
    & Atmosphere & & DUNE, HK, JUNO, ...
    \\ 
    \hline
    \hline 
    \end{tabular}
    }
    \end{sideways}
    \caption{ Summary of the most significant experimental signatures for the BSM scenarios covered in this document. Example experiments sensitive to each scenario are also provided (see references for the full list). Abbreviations: Atm.=Atmospheric, $B$=Baryon number, CC=Charged Current, CEvNS=Coherent Elastic $\nu$-Nucleus Scattering, DM=Dark Matter,  FD=Far Detector, HE=High Energy, LED=Large Extra Dimensions, ND=Near Detector, NSI=Non-Standard Interactions, SN=Supernovae, UHE=Ultra-High Energy. }
    \label{tab:big_table_prospects}
    \end{table}
    
    \clearpage% Flush page
%}

\subsection{Neutrinos as Terrestrial and Astrophysical Messengers}

\begin{figure}
    \centering
    
    \includegraphics[width=0.8\textwidth]{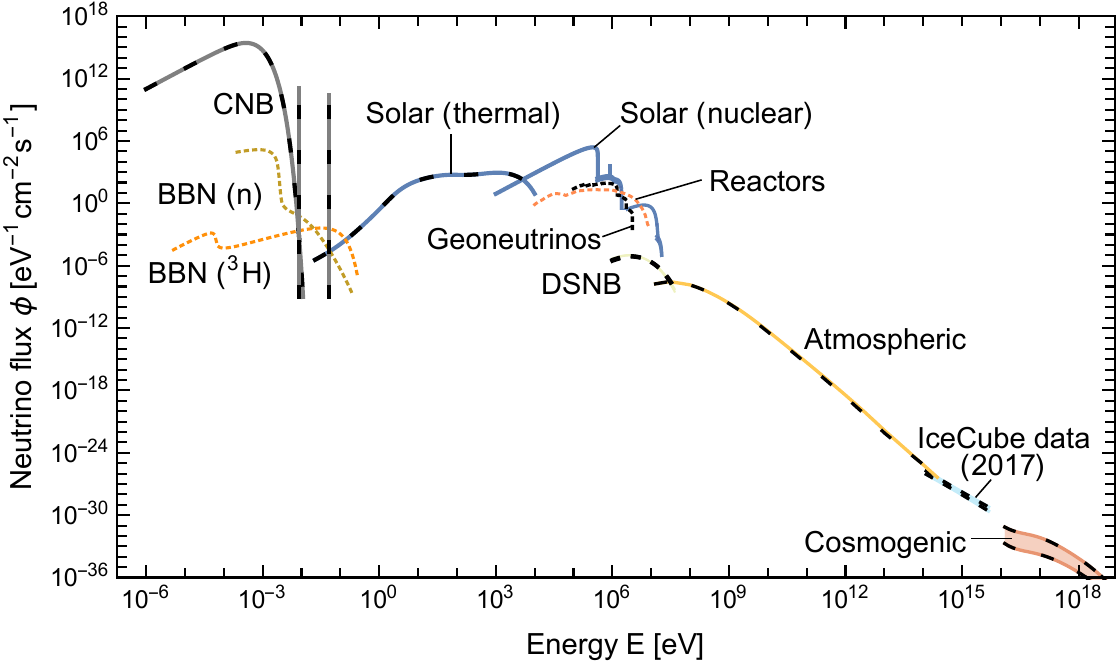}

    \caption{``Grand Unified Neutrino Spectrum" of natural neutrinos arriving at Earth  (reactor neutrinos also shown). Figure from Ref.~\cite{Vitagliano:2019yzm}. }
    \label{fig:guns}
\end{figure}

Neutrinos are produced copiously by natural sources over many orders of magnitude of energy.  The natural neutrino spectrum is summarized in Fig.~\ref{fig:guns}.  By virtue of their weak interactions, neutrinos can emerge from deep inside otherwise inaccessible sources.   They serve as astrophysical messengers,  providing unique information about the nature of the sources themselves; in conjunction with other types of emissions (photons, gravitational waves, charged particles) neutrinos provide a deeper and richer scientific story.  Furthermore, astrophysical source emission, when sufficiently well understood, enables studies of the properties of neutrinos themselves-- the discovery of neutrino oscillations in atmospheric and solar fluxes by Super-Kamiokande and the Sudbury Neutrino Observatory was a spectacular precedent for this. Cosmological studies also provide important contributions to our understanding of neutrino properties, with sensitivity to the number of neutrinos, the sum of their masses, and potential new neutrino interactions.

Detection techniques and types of information gained are highly dependent on the neutrino energy regime.    Current-generation detectors have measured neutrinos from a few hundred keV up to the PeV regime.  Expanding outside of this energy interval is unexplored territory, requiring novel technology.

In the regime from a few to a few tens of MeV, future precision measurements of solar neutrinos offer the potential to resolve uncertainties in the metallicity of our Sun, to inform our understanding of stellar evolution, to probe the interaction of neutrinos with matter in a unique environment, to search for non-standard interactions and new physics effects, and to perform precision tests of the three-flavor neutrino mixing paradigm.  Building on the ground-breaking discoveries and technological developments of the Borexino experiment, a suite of next-generation detectors is under development, which leverage novel technology to improve background discrimination in order to enhance sensitivity to these low-energy neutrinos.  SNO+ is now the largest, deepest operating liquid scintillator detector, and JUNO will come online in a few years, with a large volume and excellent sensitivity to oscillation parameters and higher-energy solar neutrino studies.  A future program can leverage multi-purpose detectors that can also target dark matter or other low energy physics;  these detectors include noble liquid detectors, or hybrid neutrino detectors, such as Theia, which seek to combine both Cherenkov and scintillation signatures for additional particle and event identification capabilities.   

A similar suite of detectors is also sensitive to antineutrinos produced within the earth -- so-called geoneutrinos.  Observations to date have been made at the KamLAND and Borexino experiments, but additional measurements are needed in order to fully interpret the results and implications in light of different models 
for heat production within the Earth.  Of particular interest are measurements of the U/Th ratio; a measurement of the flux produced from the mantle, which can be inferred using a suite of results from different geographical locations, or measured directly with an ocean-bottom detector; and the potential to detect the potassium antineutrino flux.

Moving up in energy, the next galactic core-collapse supernova is expected to be detected in photons, neutrinos, and gravitational waves and will carry a formidable amount of information on the physics of the collapse of massive stars. This multi-messenger detection will be crucial to test our conjectures on the supernova mechanism. To this end, progress is being made in order to use neutrinos to alert astronomers and gravitational wave physicists of the core collapse as soon as possible through the SNEWS~\cite{SNEWS:2020tbu} network. Neutrinos will carry clear signatures of the nature of astrophysical event, such as sharp neutrino flux cutoff that would indicate the formation of a black hole. Despite swift progress in the field in the last decade, the impact of neutrino flavor mixing in the dense supernova core remains to be assessed and it is one of the main goals for the next decade, due to its  implications on the supernova hydrodynamics, nucleosynthesis and detectable neutrino signal. In order to maximize the physics we will extract from this once-in-a-generation event, it is imperative to have neutrino experiments with high uptime and using a variety of detection technologies in order to enhance the statistics of the measurement, but also to ensure measurements of all neutrino flavors. It is expected that the next supernova burst will test the existence of new physics or place stringent constraints on it; however, for this purpose, a consistent modeling of new physics in the source is mandatory. 
The diffuse supernova neutrino background (DSNB)-- the relic neutrinos from core-collapse supernova which have occurred since the beginning of the Universe-- is another interesting target, for which first detection may be achieved in the near future by gadolinium-doped water Cherenkov experiments such as Super-Kamiokande-Gd.  The detection of the DSNB will push low-energy neutrino astronomy to extra-galactic scales and will provide complementary insight to the detection of supernova burst neutrinos on the supernova population as well as on new or exotic physics. 

Moving to the GeV regime, 
 atmospheric neutrino observation by next-generation detectors planned to operate in near future will continue to bring insights.  Information on neutrino oscillation parameters, especially $\theta_{23}$ octant, mass ordering, and $\delta_{CP}$, will enhance the knowledge from terrestrial long-baseline experiments.
Atmospheric neutrinos also have sensitivity to a range of BSM physics, including sterile neutrinos, Lorentz invariance violation, non-standard interactions, and CPT violation.
Precise measurement of the atmospheric neutrino flux is also an important topic, not only for neutrino oscillation analysis, but for other physics like baryon number violation searches, because the atmospheric neutrino signal is the primary background for these searches.
For that purpose, it is necessary to reduce the uncertainty of hadron production by primary proton interactions.
In addition, searches for prompt neutrinos from atmospheric charm production by cosmic rays in neutrino telescopes is an interesting topic.

At the highest energies,
the advent of real-time astronomy in IceCube has already led to identification of a handful of likely cosmic sources of ultra-high-energy neutrinos. The expected growing numbers of such detections as well as the larger statistics enabled by upcoming neutrino telescopes such as IceCube-Gen2 and KM3NeT will bring new understanding of particle acceleration in the sources as well as the mechanisms powering cosmic accelerators. In the next decade, we expect the emergence of radio-detection-based neutrino telescopes that will be sensitive up to the ZeV range in energy. This will bring information on cosmic accelerators at the most extreme energies, as well as information on neutrino properties and new searches for BSM physics. 

On the very low-energy end of the spectrum, 
the indirect effects of cosmic background neutrinos have  been well studied using multiple types of cosmological survey data, and these data have brought key understanding of the neutrino absolute mass scale, as well as the number of neutrino degrees of freedom.  
A long-standing, very challenging goal is to \textit{directly} measure the cosmic neutrino background detection.   In this endeavor, PTOLEMY is the only long-term project with the aim of direct cosmic neutrino background.  Cosmological data and laboratory-based data are highly complementary in this area.

\subsection{Neutrino Properties}

The existence of non-zero neutrino mass is, to date, the only laboratory-based observation of BSM physics. The disparity of mass scales between neutrinos and other fundamental particles suggests a high-energy scale for neutrino-mass generation. In models where this mechanism lies at or below the TeV scale, the physics of neutrino mass may be accompanied by complementary signatures at colliders.
In models where the scale is higher, experiments probing the nature of neutrino mass are the only feasible way of exploring this new physics. The observation of neutrinoless double beta decay would provide direct evidence that lepton number is violated, opening a path to baryogenesis via leptogenesis in the early universe.
As such, direct tests of the scale or nature of neutrino mass target some of the most central open questions in fundamental physics today.

The absolute mass scale of the neutrino is accessible through several complementary measurements, though notably not through neutrino oscillations. Laboratory probes measure the kinematics of beta decay, a field that has recently seen substantial technical and scientific advances. These measurements are complementary to astrophysical and cosmological approaches. Searches for neutrinoless double beta decay investigate the Majorana or Dirac nature of the neutrino. The next generation of these experiments at the ton-scale is prepared to begin construction early in the coming P5 period. Completion of these experiments is a continuing focus of the neutrino physics community. Pursuing the physics associated with neutrino mass was a key Science Driver in the 2014 P5 report, and the timely development and deployment of a U.S.-led ton-scale neutrinoless double beta decay experiment was a top priority item in the 2015 Nuclear-Physics Long-Range Plan, a commitment that continues today under the stewardship of the Department of Energy Office of Nuclear Physics. A rich research and development program toward beyond-ton-scale sensitivities is underway. The envisioned experiments would be sensitive to a wide range of neutrino-physics phenomena, and the technologies under development may have broad applications in particle physics and beyond. 

Other neutrino properties may be connected to extensions of the standard model, yet are not observable via oscillations.
Neutrino electromagnetic properties are of fundamental interest, and  the nascent program measuring coherent elastic neutrino-nucleus scattering (CEvNS) offers intriguing sensitivity~\cite{Abdullah:2022zue}.  Lorentz and quantum mechanical properties of neutrinos may also illuminate new physics that is inaccessible through other techniques.

\subsubsection{Neutrino-Mass Scale}
\textit{What is the absolute mass scale of the neutrino? Complementary experimental approaches currently set limits on this mass scale; when a non-zero mass is measured, will it appear in other probes in a way consistent with our current understanding?}

Direct, kinematic measurements of the neutrino-mass scale are essential to disentangle this property from the model dependence of cosmological concordances, supernova dynamics, or neutrinoless double-beta decay. The improving sensitivity of all neutrino-mass-measurement techniques raises the possibility of a fruitful disagreement between methods. A measured neutrino mass $m_{\beta}$ within the projected KATRIN sensitivity~\cite{KATRIN:2022ayy} would constrain the available model space for neutrinoless double-beta decay, and require the introduction of new physics to be reconciled with current cosmological data~\cite{Abazajian:2022ofy}. Any positive measurement should be followed up using a different experimental technique and/or a different decaying isotope. There is a strong consensus to pursue realization both of cyclotron-radiation emission spectroscopy for a next-generation tritium experiment such as Project~8~\cite{Project8:2022wqh}, and of microcalorimetry with embedded \textsuperscript{163}Ho as developed by ECHo and HOLMES~\cite{Ullom:2022kai}, to ensure this flexibility. Continued effort to identify additional isotopes for kinematic $m_{\beta}$ measurements could open up new experimental possibilities.

\subsubsection{The Nature of Neutrino Mass}
\textit{What is the mechanism that generates neutrino mass? Is the neutrino a Majorana fermion or a Dirac fermion? How can the sensitivity of neutrinoless double-beta decay searches best be improved beyond the inverted-ordering region targeted by the next generation of experiments?} 

Detection of neutrinoless double beta decay is the only known method with plausible sensitivity to the Majorana nature of the neutrino, one of the most important open questions in particle physics. Techniques aimed at its discovery have been developed in several isotopes with tens-to-hundreds~kg scale demonstrators paving the way for ton-scale and larger discovery-class experiments over the last Snowmass period. The proposed experiments use a range of detection techniques: loaded liquid scintillator, gas and liquid time projection chamber, bolometer, and solid-state detectors are all competitive approaches. 
Certain experimental challenges, notably isotope procurement and background measurement and control, are broadly shared among experimental schemes. 
The coming ton-scale generation of experiments will probe effective Majorana neutrino masses, $m_{\beta\beta}$, as small as 10\,meV in discovery mode, fully exploring the parameter space favored by the inverted mass ordering and covering a large fraction of the parameter space associated with normal neutrino mass ordering under the light neutrino exchange mechanism.

A portfolio of international experiments, representing multiple isotopes and detection technologies, will both probe this parameter space and, importantly, permit confirmation of any discovery. There is strong consensus for the goal of building at least two ton-scale experiments with U.S. leadership, multi-agency, and international support, as well as continuing participation in other programs worldwide.  

The neutrino community is pursuing a thriving R\&D program for beyond-ton-scale neutrinoless double-beta decay searches (e.g.~\cite{CUPID:2022wpt,Jones:2022moh}). This program offers exciting opportunities for upgrades that would improve the sensitivity of ton-scale experiments, and for future sensitive, ultra-low-background detectors that can access a broad physics program. Support for R\&D in multiple isotopes and with multiple technologies is recognized as vital to ensure readiness to fully explore the range of half-lives associated with normal neutrino mass ordering, or to confirm and pursue precision measurements in the case that neutrinoless double-beta decay is discovered at the ton scale.  Theoretical progress in computation of nuclear matrix elements is also critical in this endeavor~\cite{Cirigliano:2022oqy}.

\subsubsection{Neutrino Electromagnetic Properties}

\textit{Are the electromagnetic properties of the neutrino consistent with standard-model predictions arising from radiative effects?}

In the standard model, neutrinos have small charge radii induced by radiative corrections.
The predicted values of the electron and muon neutrino charge radii are less than an order of magnitude smaller than the
current experimental upper limits and can be tested in the next generation of accelerator and reactor experiments
through the observation of neutrino-electron elastic scattering and CEvNS.
Precision measurements of the neutrino charge radii would either be an important confirmation of the standard model, or would discover new physics.
The same types of experimental measurements
are also sensitive to more exotic neutrino electromagnetic properties:
magnetic moments and millicharges,
which would be certainly due to new BSM physics.
The discovery of millicharges or anomalously large neutrino magnetic moments 
would have also important implications for astrophysics and cosmology.

\subsubsection{Other Neutrino Properties}
\textit{Can probes of other neutrino properties, such as the neutrino lifetime or Lorentz-invariance violation in the neutrino sector, shed light on physics beyond the standard model?}

Since neutrinos are very elusive,
they can have very exotic properties that have not been discovered so far,
e.g., properties that violate the Lorentz and CPT symmetries
and gravitational interactions that violate the equivalence principle.  Data from a wide range of neutrino experiments have already been used to constrain many of these exotic properties.
There is no specific experimental plan aimed at investigating these properties,
but experimentalists and phenomenologists should be alert to opportunities to use new experimental data
for the exploration of all non-standard neutrino properties.

\subsection{Neutrino Interactions}
\label{sect:interactions}

A thorough understanding of neutrino cross sections in a wide range of energies is crucial for the successful execution of the entire neutrino physics program.
Support is needed for all stages of achieving this understanding, including new experimental measurements (e.g.~\cite{Alvarez-Ruso:2022ctb}), improvements to theoretical neutrino scattering models~\cite{Ruso:2022qes}, and refinements to interaction simulations.

Neutrino interactions constitute either signal or background for a variety of exciting physics measurements planned in the near and far future. The scientific reach of these efforts will ultimately depend on obtaining a detailed knowledge of the interaction physics. The complex nuclear targets used by most experiments present a particular challenge for obtaining the needed theoretical precision. A far-reaching program of new measurements is needed to inform theoretical improvements. The success of the field will also depend on support for this foundational theoretical work and its implementation in \textit{event generators}, the computer programs used to simulate neutrino scattering for experimental analyses~\cite{Campbell:2022qmc}.

There are valuable proposed cross-section measurements applicable to future programs, including new electron scattering measurements, new pion scattering measurements, measurements at long- and short-baseline experiments, and dedicated neutrino scattering measurements. These measurements provide new information to develop interaction model theory and improve its implementation in event generators.   This is another area of overlap with the nuclear physics community.

The community faces important challenges:
\begin{itemize}
    \item The exact role of and impact of the suite of new cross section measurements on oscillation physics is not yet completely elucidated.
    A dedicated exercise, overseen by oscillation experimental programs but also involving theory and external measurements, is needed to assess the benefits of those measurements, and to refine what specific measurements could be valuable.
    \item  Event generators are a key tool for neutrino experiments, as elsewhere in HEP~\cite{Campbell:2022qmc}. There are several available generators, each using different approaches, and all valuable to the community. However, improvements to these generators can take significant time to implement, and it is challenging to keep them up to date with state-of-the-art modelling. This is in part due to the necessary work being undervalued relative to other activities. The situation needs to be improved, and incentives aligned with the needs of the experimental program should be provided.  We also advocate for continued grassroots efforts to identify and resolve short-term issues and to identify how generators should interface most constructively with experiments.
    \item There will be a wealth of important experimental cross-section data in the short term, but the needs of the oscillation program may change. We endorse efforts by experimental collaborations to have data preservation plans which allow for re-analysis of the unique capabilities of experiments.
\end{itemize}

\subsubsection{Low-Energy Cross Sections}

Low-energy neutrino scattering on nuclei will be a signal or background process in a variety of exciting physics measurements in the near and far future.
The recent observations of CEvNS have spurred renewed interest in the process as a probe of fundamental properties of the neutrino, weak interactions, and nuclear properties. First-light measurements have already produced significant new constraints in these areas and the impact of future precision measurements of the total cross sections and recoil distributions is still being explored theoretically. This process will also be an irreducible background for direct dark-matter searches and a controllable background for accelerator-produced dark-matter searches at stopped-pion facilities where preservation of the timing structure of the neutrino flavors reduces the CEvNS background relative to the dark-matter signal.  

Low-energy inelastic scattering processes provide the foundational detection mechanism for solar, reactor, stopped-pion-based oscillation experiments and supernovae. In many cases, the cross sections for these interactions have never been measured, and the simulation capabilities needed to interpret future measurements are very limited. In some cases, there are nuclear physics topics interesting in their own right that can be explored in dedicated neutrino experiments.
\begin{itemize}
    \item Future planned CEvNS (COHERENT~\cite{Akimov:2022oyb}, CCM~\cite{VandeWater:2022qot}) measurements will make significant improvements in precision that expand our understanding of neutrino properties and extend the sensitivity of BSM tests and hidden-sector searches; electron-scattering experiments play a vital role in isolating modifications to the cross sections from the finite size of the nucleus.
    \item Measurements of inelastic cross sections on nuclei relevant for oscillation and supernova physics are critical to the success of these programs. While facilities and detection technologies exist to make these measurements, focused and dedicated efforts will be required to achieve the needed precision.
\end{itemize}

\subsubsection{Medium-Energy Cross Sections}

Accelerator-based short- and long-baseline neutrino oscillation experiments operate in the medium energy regime of $\sim$0.1--20~GeV. A precise extraction of oscillation parameters by these experiments requires relevant neutrino cross sections to be well modeled. Future programs recognize this and have designed highly capable suites of near detectors with new capabilities to make needed cross-section measurements. The DUNE collaboration plans to use precision detectors to make measurements at multiple off-axis angles.

Various processes, including quasi-elastic scattering, multi-nucleon knockout, resonance production, and both shallow- and deep-inelastic scattering play important roles at medium neutrino energies\cite{Workman:2022ynf,Mosel:2016cwa,SajjadAthar:2022pjt}. The neutrinos generated by accelerators and atmospheric sources have a broad energy spectrum. Multiple channels contribute to the observed event rates, and, because neutrino oscillations are energy-dependent, the cross section for each process as a function of energy needs to be well understood (with well-quantified uncertainties) to interpret measurements correctly. Rare charged- or neutral-current processes may also be important as signal or background as well, especially in searches for exotic physics. Nuclear dynamics of the neutrino target material (commonly carbon, oxygen or argon) greatly complicate the needed theoretical calculations. Furthermore, neutrino experiments also need predictions for all relevant flavors of neutrinos ($\nu_e$, $\nu_\mu$, $\nu_\tau$) and antineutrinos. Reliable energy-dependent total cross-section predictions are essential but insufficient on their own: accurate estimation of the efficiency and purity of an experimental selection may depend on various details of the composition and kinematics of exclusive final states. Accounting for missing energy (due to, e.g., undetected neutrons) when performing neutrino calorimetry is also a critical task that requires full exclusive final-state modeling. The unprecedented increases to beam exposure and detector size also enable explorations of final states in increasing detail.

Various efforts are currently underway or proposed which will provide useful results for better understanding neutrino interactions. These include indirect but highly valuable new  measurements with electrons~\cite{Ankowski:2022thw} (E12-14-012, e4$\nu$, LDMX, A1, eALBA) and pions (LArIAT, WCTE, ProtoDUNE) as well as direct neutrino cross-section measurements by long-baseline experiments (T2K, NOvA, DUNE, Hyper-K), short-baseline experiments (MicroBooNE, SBND, ICARUS) and dedicated neutrino scattering experiments (ANNIE, MINERvA, NINJA, H/D bubble chambers, FASERnu, SND@LHC, nuSTORM).

\subsubsection{High-Energy Cross Sections}

Understanding neutrino interactions at energies from the few hundreds of GeV to the TeV scale and beyond is important for studying ultra-high-energy astrophysical neutrinos.
In this energy regime, deep inelastic scattering of neutrinos off individual quarks inside nucleons is the dominant interaction mode. Cross-section calculations with small uncertainties are available up to 100~PeV. At still higher energies, theoretical predictions currently have relatively large QCD-related uncertainties. Laboratory measurements of neutrino scattering at very high energies are difficult; no neutrino cross-section data are yet available for energies above about 350~GeV, although information is available from atmospheric neutrinos~\cite{Aartsen:2017kpd,IceCube:2020rnc,Bustamante:2017xuy}. FASERnu, SND@LHC, and experiments at the Forward Physics Facility (FPF) will extend experimental reach into the TeV scale.

\section{Enabling Tools and Technology}

In this section we describe the tools and technology that enable the success of Neutrino Frontier physics, including computing and software, accelerators, and detector instrumentation.  In many cases, relevant issues and opportunities are shared with other physics Frontiers; in other cases, problems in neutrino physics require specialized tools to solve them. 

\subsection{Computing and Algorithms}

\subsubsection{The Changing Computing Landscape} 

Computing is a critical enabling technology for neutrino experiments, as it is across all of High Energy Physics, for both small~\cite{FASER:2022yqp} and large~\cite{Fleming:2022nvv} experiments. A central challenge facing all of us is that computing is undergoing a phase change: our history of using “embarrassingly parallel” resources like the Open Science Grid will not be able to continue to increase in scale to meet our needs. This change reflects an underlying dynamic beyond the control of the HEP community: the speed of individual computing cores is no longer increasing as it once did, the so-called “end of Moore’s law,” and thus the international funding agencies are no longer prioritizing investment in these types of computing resources. The good news is that there are ways past this barrier which make use of more specialized “accelerator” hardware like GPUs and FPGAs, and national and international funding agencies have been investing in computing based on these technologies. Therefore, in principle, enough “computing cycles” are expected to be available in the future to enable neutrino science. However, these ``High Performance Computing" (HPC) facilities typically take the form of a smaller number of specialized facilities (“leadership class” computers or “supercomputers”). We should anticipate a more complicated computing landscape, with different facilities optimized for different types of work. It is plausible that simulation, plot making, statistical analysis, and machine learning will all happen on different computing resources optimized for each type of problem. This challenge is not limited just to large experiments like DUNE with significant computing organizations, but will affect all experiments at all scales, so funding opportunities for computing development should be inclusive. Given the fast-paced change in these technologies, and the shared nature of the challenge across all areas of HEP, we strongly support the Computing Frontier recommendation to create a HEPAP subpanel on computing to advise and coordinate across the field at a higher frequency than Snowmass planning exercises. 

Adapting successfully to the move to specialized computing architectures will require significant new software development since we are only in the earliest days of making use of these facilities in many neutrino experiments, as highlighted by Critical Challenge 3 in the Computing Frontier report. Achieving this adaptatioon will require support for the development of both shared software solutions (Critical Challenge 2) where they are possible as well as in experiment-specific contexts. DUNE provides examples of both: it is already making use of some LHC-developed tools for data movement and workflow management, which shows the possibility for sharing solutions. However, DUNE will also have specialized needs which differ from colliders.  For example, DUNE will have a relatively small number of events compared to an LHC experiment, but each normal event will be gigabytes in size, making processing a challenge in normal circumstances. In addition, DUNE also anticipates recording supernova burst candidates approximately once per month, and each of these candidates will produce terabytes of data over the course of 100 seconds which then needs to be processed quickly to provide potential early warning to the other observatories. Preparing for these and the many other experiment-specific computing challenges will require training a new generation of physicists how to make effective use of these new computing architectures and providing a career path for them so their expertise is not lost, which is highlighted in Critical Challenge 4. The move to these computing paradigms also presents some organizational challenges: we need an access model which allows trust to flow from the experiments on to the owners of computing (as we currently have with the OSG), and experiments will need to adapt to proposal-driven, rather than fair-share, distribution of computing resources, which requires additional planning. 

\subsubsection{Common Software Tools}

Neutrino experiments' computing needs extend beyond hardware into common software challenges. These shared tools can be broadly divided into simulation packages and software frameworks. These tools are faced with two Critical Challenges as outlined in the Computing Frontier report: both the general need for continuous development of existing tools (Critical Challenge 1) and the challenge of supporting work for cross-cutting software (Critical Challenge 2). 

All areas of neutrino physics, from early design and R\&D to mature experiments focused on data analysis rely critically on a suite of simulation tools shared across the community~\cite{Banerjee:2022jgv}. GEANT4 is the de-facto standard in detector simulation, and in many experiments is a large, if not dominant, share of the total computing burden. A critical part of the challenge of moving to modern, heterogeneous computing is development work in GEANT to support these new environments. Microphysical simulation tools in neutrino physics (e.g., neutrino interaction generators, liquid noble element simulations) present a different challenge. These tools generally are not costly computationally. Instead, because of their wide variety, they present challenges of coordination and the need to develop common interfaces so they can be put into broad use. All of these tools further depend on physics development to take advantage of new theoretical work and new measurements such as test-beam experiments measuring hadron-Ar cross sections or short-baseline experiments measuring neutrino scattering cross sections. The development of neutrino interaction generators is particularly vital since it is a key part of achieving the unprecedented $1\%$ systematic-uncertainty targets of the DUNE experiment. 

Modern neutrino experiments are increasingly relying on a variety of shared software frameworks. Some of these frameworks have been developed and are supported by the national labs and serve only HEP experiments, or in some cases only neutrino experiments. Examples include ART, LArSoft, and Wire-Cell. Given the central role these frameworks play in recording and using experiment data, continued support throughout and even past the lifetime of the experiments using them is critical to enable continued use of the data to make physics measurements. Additionally, neutrino experiments are making increasing use of ``public" frameworks, especially in the area of machine learning, such as TensorFlow and PyTorch. Already, using these frameworks creates a support challenge where the frameworks integrate into experiment software stacks, since these external products typically update much faster than experiment software does. These challenges will only increase as the use of these frameworks increases, and particularly if we find we are contributing modifications back into the frameworks to better support our data formats and workflows. 

\subsubsection{Data Preservation}

Achieving DUNE’s $1\%$ systematic uncertainty target requires more than just generator development, however. Data have been taken and will soon be taken at a range of scattering experiments.  However, the critical time for the use of these data will be well after the end of those experiments,  e.g., during the later stages of the DUNE experiment when systematic uncertainties become increasingly important. It is clear from the experience of trying to use older data today (e.g., hydrogen and deuterium bubble chamber data) that we need to preserve these data and their analysis context with much greater fidelity, so that they can be continuously re-interpreted through newly developed theoretical models. The MINERvA experiment is working actively on this project now, but their effort, and those of current and near future experiments, would benefit substantially from cross-community tools and best practices. The experience of the nuclear physics community also suggests that continuous institutional support will be required to ensure the utility of and access to data after the end of the current experimental effort~\cite{Bailey:2022tdz}.  This need is highlighted in Critical Challenge 1 of the Computing Frontier report. 

\subsubsection{Machine Learning}

Neutrino experiments were early movers in the use of modern machine learning (ML) techniques since computer vision techniques like convolutional neural networks were readily adaptable to the relatively large, homogenous detectors commonly used to detect neutrinos. The use of machine learning has rapidly expanded to a wide range of different techniques applied to different types of problems, as has occurred across the HEP community~\cite{Albertsson:2018maf}. A key feature of modern deep learning, which differs from past ML efforts in HEP, is that it relies heavily on externally developed frameworks. This change presents an opportunity, since these frameworks are supported by large communities and often have good support for modern accelerated computing hardware (Critical Challenge 3), but it creates a new need to develop interfaces between these ML frameworks and the tools more commonly used in HEP~\cite{Harris:2022qtm}.  There is also strong motivation to integrate ML/AI capabilities closer to the data source at the detector.  This development work has the potential to support experiments at all scales if it is pursued as a cross-cutting effort rather than individually within experiments (Critical Challenge 2).

\subsection{Artificial Neutrino Sources}

The primary artificial neutrino sources of wide use in the Neutrino Frontier are accelerator-produced beams. Reactors have been and are planned to be used for multiple experiments, and future ideas for novel sources exist as well.

\subsubsection{Horn-Focused Neutrino Beams}
\label{sect:hornbeams}

Conventional decay-in-flight neutrino beams use high-energy protons striking a target to generate short-lived
hadrons (mainly $\pi^{\pm}$ and $K^{\pm}$) that are focused using magnetic horns before decaying into neutrinos. Selecting positively charged hadrons results in a beam that is primarily \numu, while selecting negatively charged hadrons results in a beam that is primarily \anumu. Current long-baseline experiments require hundreds of kW proton beam power to produce sufficient neutrino flux at their detectors.

\begin{figure}
    \centering
    \includegraphics[width=0.9\linewidth]{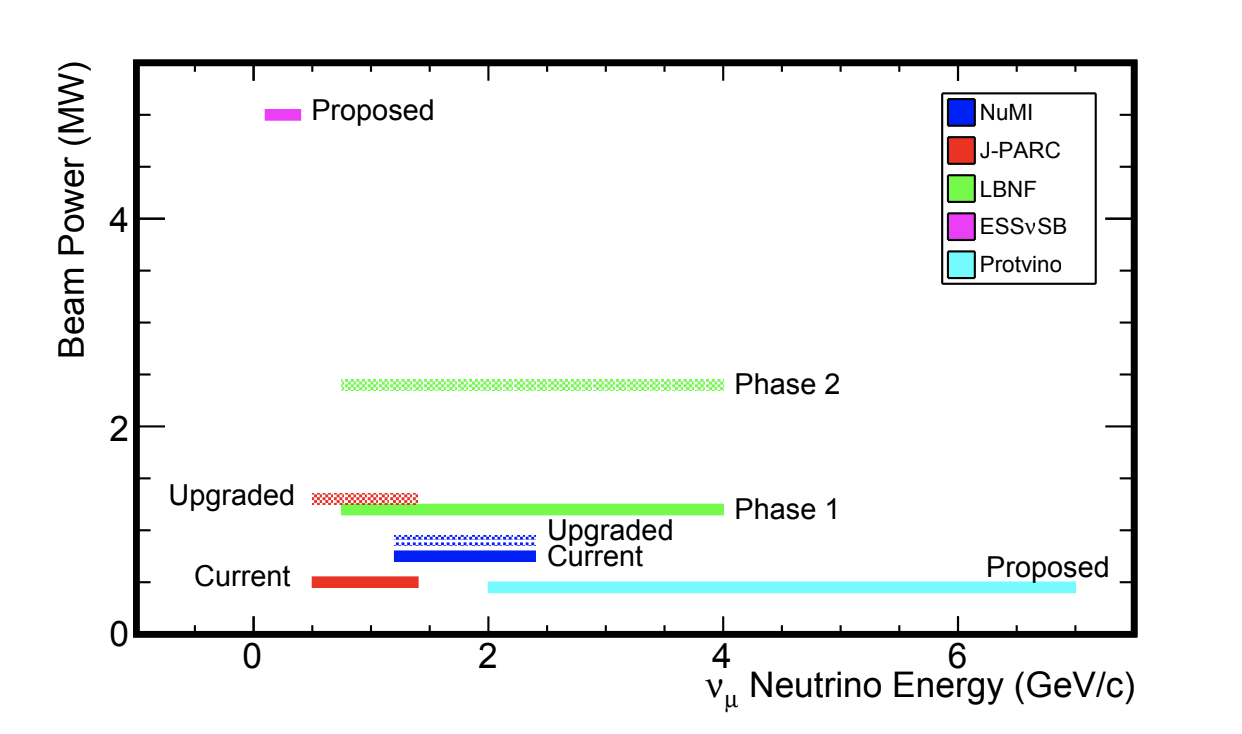}
    \caption{Neutrino beam power vs. energy for current and future long-baseline neutrino beams. The energy
range shown here is the approximate width to half-max around the beam peak energy.}
    \label{fig:beampowerplot}
\end{figure}

The current landscape for neutrino beams that result from focused hadrons includes the long- and short-baseline program at Fermilab and the J-PARC neutrino beam to T2K. Both of these laboratories will be a major international focus for long-baseline experiments over the next decade with the upgrade of the J-PARC beamline for Hyper-Kamiokande in Japan and the construction of the initial phase of
the LBNF neutrino beamline for DUNE at Fermilab. J-PARC and LBNF Phase I have planned proton beam powers of 1.3 and 1.2 MW, respectively. A beam power upgrade of the LBNF beamline from 1.2 to 2.4 MW in the second phase of DUNE will be crucial to maximizing its physics reach in a timely manner. Fig.~\ref{fig:beampowerplot} shows the beam power and energy for current and future long-baseline beams.

The Fermilab beam upgrade will be accomplished in two phases. The PIP-II project will replace the current 400 MeV linac at Fermilab with a new 800 MeV H$^-$ accelerator capable of delivering the 1.2 MW power for Phase I of DUNE. However, after this upgrade the amount of beam that can be transmitted to the Main Injector will be limited by the capacity of the 8~GeV Booster synchrotron. To reach the higher power needs of DUNE Phase II, the Booster will need to be
replaced or upgraded. The Booster replacement could be based on a continuation of the 800 MeV linac up to 2-3~GeV~\cite{Eldred:2022vxi}, which can then be followed by either a new rapid cycling synchrotron or by continuing the linac to all the way up 8 GeV~\cite{Belomestnykh:2022kal}. Modifications to  the LBNF beamline will be required to handle the higher beam power in Phase II~\cite{HurhCSS}. Some components such as the shielding, absorber, and decay pipe, are being designed for a 2.4 MW beam. However other components, such as the target and first horn, will need to be redesigned for Phase II. Significant R\&D on target materials will be essential to ensure target reliability in Phase II and allow for accurate prediction of the component lifetimes~\cite{PellemoineCSS}. Additional information as well as details on physics opportunities with Fermilab beam upgrade paths are discussed in \cite{Nagaitsev:2021xzy,Arrington:2022pon,Ainsworth:2021ahm,Toups:2022yxs,Toups:2022knq,Eldred:2022vxi}.

A design study has recently been produced on the feasibility of creating an intense neutrino super beam with a beam power of up to 5 MW using 2.5 GeV protons at the European Spallation Source~\cite{Alekou:2022emd}. There is also a proposal to send a neutrino beam from Protvino to KM3NeT/ARCA~\cite{Akindinov:2019flp}.

Near detectors and other beamline instrumentation are required to constrain flux uncertainties, which result from uncertainties in hadron production and uncertainties in the parameters of the beamline, such as horn current and position. 
Measurements of the production of hadrons in the interactions from dedicated hadron production experiments also improve our knowledge of the neutrino fluxes in accelerator-generated beams. The NA61/SHINE, EMPHATIC and DsTau/NA65 experiments are in the midst of program of measurements aimed at improving flux determination at these beamlines.  Additionally, upgrades to the Fermilab accelerator complex~\cite{Pellico:2022dju, Ainsworth:2021ahm} will enable new proton beam dump facilities at PIP II (PIP2-BD~\cite{Toups:2022yxs}) and the SBN (SBN-BD~\cite{Toups:2022knq}).

Potential future opportunities with focused beams include time-bunched sources, in which a tightly-bunched proton beam and precision timing detectors can be used to separate a wide-band neutrino beam into narrower components by selecting events based on their time of arrival at the detector~\cite{StroboscopicLOI,Angelico:2019gyi}, and optimization of beams to generate higher energy neutrinos allowing for more precise measurements of \nutau appearance.  
Another opportunity is ENUBET~\cite{Longhin:2022tkk}.

\subsubsection{Neutrinos from Stopped-Pion Sources}

Spallation neutron sources provide, as a by-product, the most intense accelerator-based sources of neutrinos in the
world, and are particularly important for a wide variety of neutrino physics measurements, including
short-baseline oscillations, neutrino interaction studies relevant for oscillation searches at both short-
and long-baseline, CEVNS and related new physics searches involving
non-standard neutrino interactions, and probes of dark matter using nearby neutrino detectors. These
physics measurements are also possible at dedicated, non-spallation-neutron facilities.

Neutrinos from stopped-pion sources have the advantage of fluxes with well-known energy, flavor and time dependence.
The dominant flavor components (in some cases dominant to the extent of very high purity) of the flux are $\nu_\mu$ with precisely known energy of 29.8~MeV from the two-body $\pi^+$ decay at rest, followed on a 2.2-$\mu$s timescale by $\nu_e$ and $\bar{\nu}_\mu$ from the three-body muon decay at rest.  More tightly pulsed proton-on-target beams are favorable, as are large and dense targets that stop a high fraction of pions produced, resulting in a well-understood decay-at-rest spectrum.  Neutrino production per power is maximized for protons of around 1.3~GeV~\cite{COHERENT:2021yvp}.  Several existing and new facilities are planned, including at the Oak Ridge National Laboratory Spallation Neutron Source (site of COHERENT~\cite{Akimov:2022oyb}), for which an update to a total power of 2.4~MW including a Second Target Station is planned~\cite{Asaadi:2022ojm}, J-PARC (site of running JSNS2~\cite{JSNS2:2021hyk} and planned JSNS2-II), and the European Spallation Source.

\subsubsection{Nuclear Reactors}

Nuclear reactors have had a central role in experimental neutrino physics ever since the discovery of these elusive particles in 1956. As neutrino sources, they have the following advantages:
\begin{itemize}
\item Intensity: about 2 × 10$^{20}$ electron antineutrinos are produced every second from a 1\,GW (thermal)
commercial reactor core. They are in fact the most intense artificial source of neutrinos.
\item Availability: there are currently over 400 nuclear reactors in operation in more than 30 countries
worldwide. The world’s nuclear power capacity continues to increase steadily, with about 55
new reactors currently under construction.
\item Flavor purity: antineutrinos are produced from the beta decays of nuclear fission products through the $n \rightarrow p + e^{-} + \bar\nu_e$ reaction. Consequently, only electron antineutrinos are emitted.
\item Predictability: The various flux anomalies non-withstanding, reactor neutrino fluxes are known in both rate and shape to within about 5\%. This fact, combined with a very well known detection cross section for inverse beta decay, allows for precision physics studies.
\item Cost-effectiveness: researchers do not typically have to bear the costs involved in designing,
building, and operating nuclear reactors, whose primary goals include power generation (commercial reactors) and neutron production (research reactors). This in turn allows for small experiments with low barrier to entry and faster timescales from design to data-taking compared to
large international projects, providing important opportunities for workforce development.
\end{itemize}

Nuclear reactors continue to be widely used as MeV-scale neutrino sources and a next-generation reactor neutrino program spanning a wide range of physics goals is under preparation. These experiments will benefit from the precise characterization of reactor neutrino fluxes made by experiments in the last
decade, which uncovered discrepancies with the models while also providing important clues about their origin. Direct neutrino measurements with greater precision, as well as in the uncharted region below the 1.8\,MeV inverse beta decay (IBD) threshold, are anticipated in the next decade that will further enhance existing constraints on reactor neutrino emission. The reactor neutrino community is also interested in pursuing ancillary nuclear physics measurements that will allow to fully resolve existing data vs. model discrepancies and that will enable improved reactor neutrino flux modeling. 

\subsubsection{Neutrinos from the LHC}
The FASERnu collaboration~\cite{FASER:2019dxq, FASER:2022hcn} recently reported the first observations of neutrinos from a
collider~\cite{FASER:2021mtu}. This measurement is likely to be the first of many neutrino measurements at the Large Hadron Collider (LHC). Physics goals for LHC-based neutrino experiments include cross section measurements of all neutrino flavors at previously unmeasured neutrino energies, searches for BSM physics, and neutrino flux measurements as novel constraints to LHC event generators~\cite{Bai:2022jcs,Anchordoqui:2021ghd,Feng:2022inv}.

\subsubsection{Novel Neutrino Sources}

Muon-based neutrino sources offer advantages over traditional hadron-focused
neutrino beams in that they contain equal parts muon neutrinos and electron antineutrinos and have a well known energy spectrum and flux. Several facilities have been proposed to produce and make use of neutrinos from stored muons. Because they share common challenges associated with producing, accelerating, and storing muon beams, there are many overlaps in R\&D efforts towards these facilities. Addressing these challenges was not identified as a priority in the 2014 P5 report, but was set
as a high priority in the 2020 update of the European Strategy for Particle Physics. Continued
development of these concepts would greatly benefit from an endorsement in the upcoming P5 process. Three stages in the development of stored muon beams can be identified.  The first stage is the nuSTORM facility~\cite{nuSTORM:2022div}, storing 1-6\,GeV muons and producing beams of muon and electron neutrinos spanning the
0.5 - 5\,GeV energy range without any muon cooling. Its primary physics goal is measurement of electron and muon neutrino cross
sections at the few percent level, but it would also provide measurements of various nuclear effects,
searches for sterile neutrinos, and serve as a test facility for the development of a future neutrino
factory and/or muon collider. Both Fermilab and CERN would be capable of hosting nuSTORM. 

A step beyond nuSTORM would be a  neutrino factory storing muons in the 5-50\ GeV energy range with muon cooling~\cite{Bogacz:2022xsj}. A key feature of a neutrino factory is the ability to produce a well-understood high-energy neutrino beam. As such, it would provide high-precision measurements of oscillations to tau neutrinos, opening up new tests of
the three-flavor mixing paradigm that are not currently possible. It would have greater precision in the measurement of $\delta_{CP}$ than DUNE or Hyper-K and would facilitate a wide variety of searches for BSM at both the near and far detector. The final stage then would be TeV-scale muon collider with full 6D muon cooling. If a muon collider is part of the portfolio of future colliders within the energy frontier, then both NuSTORM and a neutrino factory would be useful R\&D platforms.

Cyclotrons represent another novel artificial source of neutrinos~\cite{Winklehner:2022rdu,Alonso:2022mup}. The IsoDAR experiment~\cite{Alonso:2022uar} will utilize a high power (600 kW) cyclotron to produce a 60 MeV proton beam~\cite{Winklehner:2021qzp}. Beam interactions with a beryllium target create an intense source of neutrons which subsequently enter a surrounding  $^{7}$Li sleeve, thermalize, and then capture on $^{7}$Li to create the high-$Q$ beta-decay isotope $^{8}$Li. The resulting beta decay ($\tau_{1/2}=839$~ms) produces a high-energy, single-isotope electron-antineutrino source (mean antineutrino energy of 6.4~MeV) with $1.15 \times 10^{23}$ $\bar{\nu}_e$ expected in four years of livetime~\cite{Alonso:2021kyu}). While IsoDAR represents a unique source that can be paired with a number of existing and future large underground free-proton-based (e.g. H$_2$O, CH$_2$) detectors, the current plan is to pair IsoDAR with the 2.3~kton liquid scintillator detector at Yemilab in South Korea. The resulting detected 1.7 $\times 10^6$ IBD and 7000~$\bar{\nu}_e + e^-$ elastic events in four years of running will provide leading sensitivity to electron-flavor disappearance, non-standard neutrino interactions, and exotic particle production, among other physics measurements~\cite{Hostert:2022ntu,Alonso:2021kyu}. The IsoDAR@Yemilab cavern excavation has recently been completed and the experiment is now at an advanced R\&D stage.  Cyclotron-based sources have additional broader impacts~\cite{Alonso:2022roj}.

\subsection{Detectors}

Neutrino physics spans an enormous range of energies and scales: from detection of low-energy cosmic neutrinos, to keV-scale recoils in coherent neutrino scattering, to MeV-scale solar, reactor, and neutrinoless double-beta decay events, to GeV and TeV-scale detection of neutrinos from accelerators and the atmosphere, and beyond with cosmic sources in the PeV–ZeV range. While any particular
experiment tends to focus on just one or two detection approaches, the great breadth of neutrino physics means that there is an equally broad spectrum of detection technologies and methodologies.

At any one time, there are a dozen or more medium- to large-scale neutrino detectors operating worldwide, several more in the design or construction phases, and many future detectors planned. Beyond this are the diverse set of small- to medium-scale prototypes distributed across universities and labs. The focus in this section of the report is on new technologies and approaches that will enable future neutrino detectors, and thus experiments that are already built and running, or are under construction or for which technical designs exist are not discussed in great detail.

While there are many exciting detectors and enabling technologies for future neutrino experiments (many of which are described in detail in the NF10 report and associated white papers), there are a few ideas that have had a particularly large community interest: 
\begin{itemize}
    \item {\bf \underline{Broadening the Noble Liquid and Gas Physics Program:}} Many ideas are being pursued for improving liquid and gaseous TPCs, including new charge readout technologies, background reduction and analysis techniques for low-energy physics, the use of underground argon for low-background physics~\cite{Back:2022maq}, and the use of various dopants (xenon, photo-ionizing dopants) to increase photon or charge yields, or light traps like the ARAPUCAs for improved light detection.  Many of these ideas may enable a broader physics program than can be done with existing detector designs.
    \item {\bf \underline{Pursuit of hybrid Cherenkov/scintillation Detectors:}} Many different technologies are being developed for these, including water-based liquid scintillator, slow fluors, fast timing with large area picosecond photodetectors (LAPPDs) and other devices, and spectral photon sorting with dichroicons.  At very large scales like the proposed Theia detector, these could have very broad physics programs.
     \item {\bf \underline{Optimizing low-threshold neutrino detectors:}} The work here aims to expand the ever-growing CEvNS  program and to fully exploit the physics reach of CEvNS in the next decade.  It includes not just lowering energy thresholds but improving background rejection techniques, understanding detector responses at the eV-scale, and moving toward larger detector masses. Enabling technologies have many synergies with direct neutrino mass measurements and recoil-imaging directional dark matter detectors.
    \item {\bf \underline{Developing Technologies for Neutrino Detection at the PeV Scale and Beyond:}} Observations of high-energy neutrinos at large-neutrino telescopes has provided a wealth of physics and multi-messenger astrophysics, and new opportunities for studying neutrino interactions at LHC, including possibly tagging the production vertex, are particularly exciting.  Technologies for $\nu$ telescopes include radar echo detection, Askaryan effect detection, and ever-larger scale optical detection, while at the LHC they include picosecond timing synchronizations, intelligent triggering, and high-resolution tracking.
    
    \item {\bf \underline{Co-development of neutrino and dark-matter detectors:}} Both the development of new noble liquid and gas detectors and low-threshold detectors of various technologies lend themselves to development of  enabling technologies that are useful for both the Neutrino Frontier and the Cosmic Frontier.
\end{itemize}

\paragraph{Liquid Noble and Gas Detectors}

Since the last Snowmass, there has been enormous progress and diversity in detectors that use either noble liquids or gases as their target material. Such detectors have been used for dark matter searches by experiments like DarkSide, DEAP, and LZ, for neutrinoless double-beta decay with experiments like EXO-200, and as part of both the short-baseline program at FNAL with MicroBooNE, SBND, and ICARUS, and the DUNE experiment at SURF.  There are many new ideas for doing more with these detectors, from scaling to larger mass (e.g., nEXO~\cite{Albert:2017hjq}), to improving reconstruction capabilities by using pixelated charge readout in LArTPCs (e.g., LArPix, QPix) or moving to high-pressure (e.g. ND-GAr~\cite{DUNE:2022yni}, NEXT) or atmospheric-pressure (e.g. CYGNUS) TPCs, to lowering thresholds by using underground sources of LAr.  There is also interest in new approaches to detecting scintillation light in these detectors, including pixelated SiPM arrays, photo-ionizing dopants, or by increasing light coverage with ARAPUCA light-traps deployed on TPC cathodes using power-over-fiber. A common theme in the LArTPC community is the pursuit of a broader physics program than currently planned, including low-energy solar neutrinos or neutrinoless double-beta decay~\cite{Caratelli:2022llt,Para:2022gju}, much of which would be leveraged by underground or low-background argon~\cite{Avasthi:2022tjr}.  

    Thus some of the highest technical priorities in this area are:
    \begin{itemize}
    \item Larger-scale production of underground argon sources, or ways to remove $^{39}$Ar and $^{42}$Ar.
    \item Development of pixelated charge readout TPCs.
    \item Development of pixelated light readout for TPCs.
    \item Investigations of new ways for photon detection in liquid noble detectors, including photo-ionizing dopants.
    \item Advanced triggering schemes at low-energies, including machine learning techniques.
    \end{itemize}

\paragraph{Photon-Based Detectors} Neutrino Detectors that use photons as their primary carrier of neutrino interaction information have a very successful history in neutrino physics. They include Cherenkov detectors like IMB, Kamiokande, SNO, and Super-Kamiokande, and scintillation detectors like KamLAND, Double CHOOZ, Daya Bay, RENO, BOREXINO, NOVA, PROSPECT~\cite{Andriamirado:2022psq}, and SNO+. Of particular interest in the past decade or so are {\it hybrid} Cherenkov/ scintillation detectors, which can detect and discriminate between Cherenov and scintillation photons (``chertons'' and ``scintons'') in the same detector, thus allowing a very broad neutrino physics program.  Enabling technologies that would allow this have been and continue to be developed over the past decade~\cite{Klein:2022tqr}, including new materials like water-based liquid scintillator and slow fluors, new and faster devices like LAPPDs, and spectral photon sorting with devices like dichroicons.  These new technologies, employed either separately or in concert, make large-scale hybrid detectors a real possibility.  The most developed of these ideas is the proposed Theia experiment, which could sit in the LBNF beam at SURF.  

In addition to hybrid Cherenkov/scintillation detectors, new ideas for segemented detectors have also been developed.  The most developed of these is LiquidO, which would use scintillator with short scattering lengths and an array of fiber optics to create a ``self-segmented'' detector that could allow precision track reconstruction at both high and low energies.  The SLIPS idea, which removes physical segmentation by floating a scintillator volume, could allow very low-background experiments by eliminating radioactive sources in the volume.

 Looking even further ahead, some of the highest technical priorities seen in this area, which could be explored between this Snowmass and the next, are:
    \begin{itemize}
        \item Lower-cost, large-area high-quantum efficiency $\sim$ 40\% photon sensors.
        \item Lower-cost, fast timing ($\sim$ 100 ps) photon detectors.
        \item Dichroic filters that can be deposited on non-flat surfaces and with sharper cut-on/cut-off curves, even at high incidence angles.
        \item Narrow-band fluors for liquid scintillators.
        \item High-yield scintillators with attenuation lengths $>$ 40~m.
        \item High-yield ``slow'' ($\sim 10$ ns risetime or longer) fluors.
        \item Low-background fibre optics.
        \item New approaches to radiologic background. reductions, beyond levels seen in Borexino.
    \end{itemize}
    
    \paragraph{Low-Threshold Neutrino Detectors}

The development of low-threshold neutrino detectors with eV-scale resolution has become a priority in neutrino physics during the last decade since it has opened up new portals for the study of neutrino properties and the search for new physics. The COHERENT program has pioneered the use of detector technology which enabled the first observation of CEvNS and is driving since then a blooming new research field under U.S. leadership. This technological breakthrough triggered R\&D activities worldwide of low-threshold neutrino detectors based on a wide range of technologies (e.g. ~\cite{Abdullah:2022zue,Alfonso:2022meh,OHare:2022jnx,Alfonso-Pita:2022akn}). A broad and complementary CEvNS program based on small-scale experimental projects will enable precision measurements and pave the way for applications. The community will profit from multiple technological synergies with dark matter search measurements and a variety of proposed approaches. To achieve the technological goals the following challenges have to be addressed:  
\begin{itemize}
    \item Improve detector thresholds towards the eV scale. 
    \item  Develop advanced techniques for suppressing backgrounds, including the community-wide observed low-energy excess. 
    \item Establish multiplexing techniques to scale up active detector mass. 
    \item Understand the detector response at the eV scale.
    \item Increase level of automatization for applications in science, industry, and for society. 
\end{itemize}
Exploiting the strong technological interconnections with direct dark matter searches and neutrinoless double-beta decay is essential and of mutual interest for both communities. The CEvNS community will profit from the ongoing R\&D efforts on low-threshold directional recoil detectors for DM searches~\cite{OHare:2022jnx}.  Multi-ton dark matter detectors will play a crucial role for the measurements of solar and supernova neutrinos in the next decade~\cite{Aalbers:2022dzr}.  

\paragraph{High-Energy and Ultra-High-Energy Neutrino Detectors}

    Detection of neutrinos at the TeV scale and beyond~\cite{Ackermann:2022rqc} has been pioneered by big neutrino telescopes like IceCube and KM3NeT.  The physics and astrophysics and multi-messenger possibilities from these detectors is remarkably broad, and these detectors have been exceptionally successful.  Future plans for these detectors focus on moving toward even higher energies, into the EeV and ZeV regimes, which require scales going well beyond km$^3$ or new technologies, exploiting the Askarayan effect or radar echoes off ionization trails.  In many cases, the enabling ``technology'' for these telescopes is a piece of geography: polar or Greenland ice sheets, mountain ranges that can be used as targets, etc.  At the same time, there is a new opportunity at the LHC with the FPF~\cite{Anchordoqui:2021ghd,Feng:2022inv} to detect neutrinos produced in collisions, perhaps with the possibility of tagging the neutrino production vertex in a collider detector.  
    
    High priorities over the next several years for these detectors are:
\begin{itemize}
    \item  Develop further the ability for sensitive radio detection of neutrinos interacting in ice or the atmosphere.
    \item  Demonstrate at larger scales the detection of neutrinos via radar echos off ionization cascades.
    \item  Create low-cost ways of scaling to ever-larger telescopes sizes.
    \item  Create intelligent triggers for background rejection at the FPF.
    \item  Create larger-scale high-resolution tracking options for FPF neutrino events.
\end{itemize}

\subsection{Facilities}

Key to success of the experimental programs underway and planned in the Neutrino Frontier is the availability of suitable facilities.  These have been primarily provided at national laboratories, with some experiments hosted by universities. Of special importance for the Neutrino Frontier are underground facilities.  Because neutrino events are typically of low rate, cosmic rays are often a significant background, and overburden for cosmic ray shielding is either absolutely required or highly desirable, even in the case of pulsed-beam experiments for which background can be reduced by timing.  Large, multipurpose neutrino experiments require very large cavities and have specific requirements depending on the detector type.  Notably, DUNE has special requirements to accommodate 40 kton of liquid argon.  The Underground Facilities Frontier report provides more details~\cite{Bolton:2022pgb}.  Additional ideas for new facilities have been proposed as well~\cite{Monreal:2022crn}.

\section{Applications and Community Engagement}

\subsection{Applications}

In addition to the pursuit of fundamental knowledge, investment in scientific discovery is also motivated by the development of technologies and the training of a skilled workforce that can benefit society in other ways. In the case of Neutrino Physics, there are many instances of this dynamic
playing out. But somewhat surprisingly, given the inherent difficulty of neutrino detection, the direct application of neutrinos to advance other fields of research or solve societal problems is also a possibility. Examples of direct neutrino applications include monitoring of nuclear reactors for safeguards and non-proliferation~\cite{Bernstein:2019hix} and the probing the Earth’s interior via its neutrino emissions~\cite{Araki:2005qa} and tomography~\cite{Winter:2006vg,Donini:2018tsg}. There are strong synergies between the neutrino physics topics of community interest developed through this Snowmass process and direct neutrino applications. These synergies can take the form of overlapping needs in terms of technology development, workforce capabilities, facilities, and underlying scientific knowledge~\cite{bib:mutualbenefit:2021loi,Akindele:2022dpr,Akindele:2021sbh}. Specific examples of synergies highlighted during Snowmass include the following:

\begin{itemize}
\item Improved knowledge of the reactor neutrino emissions (flux and spectrum) is needed for reactor monitoring applications and would enable reactors neutrinos to continue to be a tool for discovery~\cite{bib:iaea,Romano:2022spd,Akindele:2022dpr}.
\item Neutrino detectors using the inverse beta decay or CEvNS channels with improved background suppression operating close to reactors can help to address neutrino anomalies, probe BSM physics, and provide versatile monitoring capabilities~\cite{Akindele:2022dpr}.
\item Technologies that could reduce the cost and improve the performance of large inverse beta decay neutrino detectors could provide reactor monitoring at larger distances and address a wide range of neutrino physics topics~\cite{DUNE:2022jhf,Theia:2022uyh}.
\item Projects at the intersection of neutrino physics and neutrino applications can provide broad training opportunities for community members while also opening new career pathways.
\end{itemize}

The fact that technologies and workforce capabilities developed by the field of neutrino physics can have benefits in many other areas of science and society provides an additional motivation for societal investment.
To further advance our field in this respect, it is important that community members have such possibilities in mind and actively seek opportunities to highlight the broader impact of their research. It is also critical to note that the full impact of neutrino applications can only be realized through early and attentive engagement with experts from potential end-user communities~\cite{Akindele:2021sbh}.
Given the strong synergies between neutrino physics and neutrino applications noted above, it is apparent that stakeholders from both neutrino and end-user communities would benefit from further coordination. This could take the form of joint investment in detector R\&D efforts, measurements to improve our knowledge of reactor emissions, and/or reactor-based experiments and demonstrations.

\subsection{Community Engagement}

Given that neutrino physics represents a major component of the U.S. program, there are opportunities for leadership in addressing societal issues. Additionally, the widely varying scope of activities within the Neutrino Frontier requires significant collaboration across disciplines and sectors. Commitment to community engagement, including discourse with other scientific disciplines, a diverse and welcoming environment within HEP, and education and outreach, facilitates these necessary collaborations. It is therefore critical for the community to address issues of diversity, equity and inclusion. A work environment that is welcoming to all is intrinsically desirable, conducive to scientific productivity, and necessary to attract and retain the people at the heart of our program. A strong outreach and education component to the program is also crucial for attracting the best talent to the particle physics community as well as for communicating the impact of our  neutrino physics.  The Neutrino Frontier facilities have a significant impact on the surrounding communities, so it is important to be good citizens of those communities. 

While DEI and community engagement is the responsibility of everyone in the field, host laboratories for large experiments have important leadership roles to play. 

The fact that this section comes late in this report does \textit{not} mean that it is an afterthought.  We emphasize the overarching importance of community engagement for the overall success of neutrino physics and HEP as a whole.  This is the first Snowmass in which Community Engagement has been identified as a Frontier. Challenges have arisen, mirroring those that many in the field experience in their day-to-day work and at their home institutions, as participation in the Community Engagement Frontier is often perceived as in conflict with or competing with effort in the scientific Frontiers. As a field we must find a way to reinforce the concept that Community Engagement is a necessary part of our work, rather than a volunteer effort that a few people do on the side. A cohesive, HEP-wide strategic plan that provides support, rewards, and career development paths for this work is urgently needed~\cite{Barzi:2022uki}.  

\section{Long-Term Outlook}

If the three-flavor paradigm is sufficient to fully describe the neutrino sector, it will take approximately the next two decades to fill in the main features of the picture.  After that, precision measurements will be needed to continue to test the robustness of the paradigm.    However, Nature may very well have surprises for us, and we need a program that will allow us to explore broadly and flexibly, if one of the current anomalies persists in pointing the way to new physics, or if perhaps some entirely new BSM signature makes an appearance.  It is difficult to determine detailed road-map for neutrino physics more than two decades out.  Nevertheless, we need to prepare for potential opportunities in a post-DUNE era with a broad and robust program of instrumentation development.  We describe in this section some longer-term prospects that align with the philosophy of broad exploration and readiness to pursue higher precision and new directions should the data lead us there.  

Future neutrino facilities, by design, should offer a wide range of opportunities to explore BSM physics within and outside the neutrino sector, without interfering with the measurement of neutrino oscillation parameters. Thus, it will be vital to significantly enhance the potential of these facilities by keeping BSM searches as a high priority in the design and optimization of future experiments. These efforts must include not only detector capabilities that will strengthen the signal efficiencies, but also new beamline designs that could enhance the signal-to-background ratio, in particular for backgrounds from neutrino interactions, in order to ensure the co-existence of BSM physics with precision neutrino property measurements.   In addition, in order to pursue BSM physics more effectively, enhanced numerical tools will be needed for accurate signal simulation and to understand the expected backgrounds precisely. This will require a close collaboration with other particle physics frontiers and the nuclear physics community.

The rich and diverse program that is underway or projected for the upcoming P5 period promises to provide a comprehensive understanding of the short-baseline neutrino anomalies. This will play a crucial role in the long-term neutrino program, as the phenomenology behind the anomalies will have an increasing impact in the interpretation of future neutrino measurements as their precision improves. Irrespective of what is discovered, be it new physics or conventional explanations for the anomalies, the community will be compelled to explore, develop, and support opportunities for incorporation of that understanding in the interpretation of future results, or else to develop a program of new dedicated measurements at those facilities aiming to characterize non-standard phenomena with high precision. Additionally, the upcoming P5 period provides opportunities for the development of new synergies across topics within multiple frontiers (colliders, dark matter, astrophysics, cosmology, instrumentation) that can be explored in the following decade and beyond.

The motivation for astrophysical neutrino detection extends to the indefinite future -- neutrino messengers from the sky will always have something to tell us.
Large-scale, multi-purpose detectors are likely to dominate the future for both low- and high-energy neutrino astrophysics and will guarantee an unprecedented view of natural sources. Detectors such as the proposed DARWIN, RES-NOVA, and Theia can address multiple topics such as searches for neutrinoless double-beta decay and dark matter, as well as astrophysical neutrino detection. Neutrino telescopes using optical and radio detection techniques, such as IceCube-Gen2, GRAND, and POEMMA will probe neutrino properties at energies not accessible in the laboratory and will enable the study of the most extreme systems in our Universe. At the same time, urgent efforts are required to model (non-)standard physics in astrophysical sources to take advantage of the upcoming plethora of multi-messenger datasets. Future neutrino detectors that look beyond DUNE will allow precision probes of the three-flavor mixing model, and will also the broaden the physics programs of large detectors.  One example would be a very large LArTPC filled with clean, underground argon, with charge- and light-sensitive pixels, or doped with photo-ionizing materials, to do a broad range of low-energy physics. Loaded with enriched LXe, such a detector could also look for neutrinoless double-beta decay with sensitivity below the inverted ordering region. These next-generation LAr detector ideas go by different names, such as  “SLoMo”, “SoLAr”, and “LArXe.” Another idea is the proposed Theia detector, which is a hybrid Cherenkov/scintillation detector that could do precision measurements of very low-energy solar neutrinos, diffuse supernova neutrino detection, perform searches for sterile neutrinos, and also push well beyond DUNE in precision tests of the three-flavor mixing model (including, for example, studies of the second oscillation maximum).  On the low-threshold side, much will depend on what the next generation of CEvNS detectors will see, but it is clear that detectors with (sub-)eV thresholds will bring reactor, accelerator, and solar neutrino experiments into a higher-precision era.
Should a signal of new physics be seen in the upcoming generation of experiments, it is anticipated that a push for higher-statistics detectors with even more precise understanding will be needed.

In the long term, opportunities exist to use current artificial neutrino sources to pursue new physics goals, such as optimizing focused-hadron sources for tau neutrino oscillation. There are also many exciting opportunities with new sources capable of producing more intense and/or precisely characterized beams, which can enable improved measurements of neutrino properties and searches for beyond-the-Standard-Model physics. These include compact cyclotron electron anti-neutrino sources and muon-based sources, which could ultimately result in a neutrino factory, which would have significant synergies with a muon collider program.

\section{Conclusion}

In summary, the U.S. neutrino program is poised for an exciting future.  The path is clear for the next P5 period and the decade beyond it.  DUNE is the flagship international experiment that will fill in the gaps of the three-flavor picture and explore new BSM territory, as well as be open to astrophysical signal opportunities from the skies.   Beyond DUNE,  the neutrino community has sent a clear message that a broad neutrino program across scales, connecting to other Frontiers and fields, is highly desirable for maximizing science output for the field.   Theoretical efforts are a key component of the program, and development of new technologies is critical for the long-term future.  As a final point, physics is done by humans.  We will not fully succeed in our science without making diversity, equity, and inclusion efforts an integrated, high-priority component of our work.

\newpage

\appendix
\section{Data on Experimental Programs}

\begin{longtable}{L{0.2\textwidth}L{0.2\textwidth}L{0.07\textwidth}L{0.47\textwidth}L{0.06\textwidth}}
  \caption{Table for Neutrino Frontier experiments.  Information in this table was provided by the collaborations.}\label{tab:experiments}\\
  \hline
  Experiment name&Status&Year&Physics goal&Ref.\\
  \hline
  \endfirsthead
  \hline
  Experiment name&Status&Year&Physics goal&Ref.\\
  \hline
  \endhead
  \hline
  \multicolumn{4}{r}{continued on the next page}\\
  \hline
  \endfoot
  \hline
  \endlastfoot
  \input{table.tex}
%\end{sideways}
\end{longtable}

\section*{Acknowledgements}

We are grateful for feedback from
Brian Batell,
Doug Bryman,
Thomas Chen,
Aaron Chou,
Brajesh Choudhary,
Janet Conrad,
Hooman Davoudiasl,
Dmitri Denisov,
Angelo Di Canto,
Daniel Elvira,
Jonathan Feng,
David Hertzog,
Tiffany Himmel,
Teppei Katori,
Wes Ketchum,
Yury Kolomensky,
Chris Marshall,
Andrew Mastbaum,
Kaixuan Ni,
Mark Palmer,
Frank Petriello, 
Jen Raaf,
Mary Hall Reno,
Vladimir Shiltsev, 
Michael Troxel, 
Matt Toups, 
Joseph Zennamo, and
Bob Zwaska, as well as to the many community members who participated in the workshops and the Community Summer Study, and who provided feedback on the report.

\input{common/final}

\end{document}

%% file: sections/author_list.tex
% DO NOT EDIT THIS FILE.  It is generated from a spreadsheet by a script.
% If you edit this file you will be sad when your edits are overwritten.
\author{\textbf{Frontier Conveners:} Patrick~Huber}\affiliation{Center for Neutrino Physics, Physics Department, Virginia Tech, Blacksburg, VA, 24061, USA}
\author{Kate~Scholberg}\affiliation{Department of Physics, Duke University, Durham, NC, 27708, USA}
\author{Elizabeth~Worcester\vspace{0.1in}}\affiliation{Department of Physics, Brookhaven National Laboratory, Upton, NY, 11973, USA}
\author{\\ \textbf{Topical Group Conveners:} Jonathan~Asaadi}\affiliation{Department of Physics, University of Texas at Arlington, Arlington, TX, 76019, USA}
\author{A.~Baha~Balantekin}\affiliation{Department of Physics, University of Wisconsin, Madison, Madison, WI, 53706, USA}
\author{Nathaniel~Bowden}\affiliation{Lawrence Livermore National Laboratory, Livermore, CA, 94550, USA}
\author{Pilar~Coloma}\affiliation{Instituto de Fisica Teorica UAM-CSIC, Campus de Cantoblanco, Madrid, 28049, Spain}
\author{Peter~B.~Denton}\affiliation{High Energy Theory Group,  Physics Department, Brookhaven National Laboratory, Upton, NY, 11973, USA}
\author{André~de~Gouvêa}\affiliation{Physics \& Astronomy Department, Northwestern University, Evanston, IL, 60208-3112, USA}
\author{Laura~Fields}\affiliation{Department of Physics and Astronomy, University of Notre Dame, IN, 46556, USA}
\author{Megan~Friend}\affiliation{High Energy Accelerator Research Organization (KEK), Tsukuba, Ibaraki, 305-0801, Japan}
\author{Steven~Gardiner}\affiliation{Physics Simulation Department, Fermi National Accelerator Laboratory, Batavia, IL, 60510, USA}
\author{Carlo~Giunti}\affiliation{Istituto Nazionale di Fisica Nucleare (INFN), Torino, Italy}
\author{Julieta~Gruszko}\affiliation{Department of Physics and Astronomy, University of North Carolina at Chapel Hill, Chapel Hill, NC, 27599, USA}\affiliation{Triangle Universities Nuclear Laboratory, Durham, NC, 27708, USA}
\author{Benjamin~J.P.~Jones}\affiliation{Department of Physics, University of Texas at Arlington, Arlington, TX, 76019, USA}
\author{Georgia~Karagiorgi}\affiliation{Department of Physics, Columbia University, New York, NY, 10027, USA}
\author{Lisa~Kaufman}\affiliation{SLAC National Accelerator Laboratory, Menlo Park, CA, 94025, USA}
\author{Joshua~R.~Klein}\affiliation{Department of Physics and Astronomy, University of Pennsylvania, Philadelphia, PA, 19104, USA}
\author{Lisa~W.~Koerner}\affiliation{Department of Physics, University of Houston, Houston, TX, 77204, USA}
\author{Yusuke~Koshio}\affiliation{Department of Physics, Okayama University, Okayama, Okayama, 7000803, Japan}
\author{Jonathan~M.~Link}\affiliation{Center for Neutrino Physics, Physics Department, Virginia Tech, Blacksburg, VA, 24061, USA}
\author{Bryce~R.~Littlejohn}\affiliation{Department of Physics, Illinois Institute of Technology, Chicago, IL, 60616, USA}
\author{Ana ~A.~Machado}\affiliation{Department of Cosmic Rays and Chronology, University of Campinas, Campinas, SP, 13083-859, Brazil}
\author{Pedro~A.N.~Machado}\affiliation{Theory Division, Fermi National Accelerator Laboratory, Batavia, IL, 60510, USA}
\author{Kendall~Mahn}\affiliation{Department of Physics and Astronomy, Michigan State University, East Lansing, MI, 48824, USA}
\author{Alysia~D.~Marino}\affiliation{Department of Physics, University of Colorado Boulder, Boulder, CO, 80309, USA}
\author{Mark~D.~Messier}\affiliation{Department of Physics, Indiana University, Bloomington, IN, 47405, USA}
\author{Irina~Mocioiu}\affiliation{Department of Physics, Pennsylvania State University, University Park, PA, 16802, USA}
\author{Jason~Newby}\affiliation{Physics Division, Oak Ridge National Laboratory, Oak Ridge, TN, 37831, USA}
\author{Erin~O'Sullivan}\affiliation{Department of Physics and Astronomy, Uppsala University, Uppsala, 75120, Sweden}
\author{Juan~Pedro~Ochoa-Ricoux}\affiliation{Department of Physics and Astronomy, University of California at Irvine, Irvine, CA, 92697, USA}
\author{Gabriel~D.~Orebi Gann}\affiliation{Department of Physics, University of California at Berkeley, Berkeley, CA, 94549, USA}\affiliation{Nuclear Science Division, Lawrence Berkeley National Laboratory, Berkeley, CA, 94549, USA}
\author{Diana~S.~Parno}\affiliation{Department of Physics, Carnegie Mellon University, Pittsburgh, PA, 15217, USA}
\author{Saori~Pastore}\affiliation{Department of Physics and the McDonnell Center for the Space Sciences, Washington University in St Louis, St Louis, MO, 63130, USA}
\author{David~W.~Schmitz}\affiliation{Department of Physics, University of Chicago, Chicago, IL, 60637, USA}
\author{Ian~M.~Shoemaker}\affiliation{Center for Neutrino Physics, Physics Department, Virginia Tech, Blacksburg, VA, 24061, USA}
\author{Alexandre~Sousa}\affiliation{Department of Physics, University of Cincinnati, Cincinnati, OH, 45221, USA}
\author{Joshua~Spitz}\affiliation{Department of Physics, University of Michigan, Ann Arbor, MI, 48108, USA}
\author{Raimund~Strauss}\affiliation{Department of Physics, Technical University Munich, Garching, 85748, Germany}
\author{Louis~E.~Strigari}\affiliation{Department of Physics and Astronomy, Texas A\&M University, College Station, TX, 77843, USA}
\author{Irene~Tamborra}\affiliation{Niels Bohr Institute, University of Copenhagen, Copenhagen, 2100, Denmark}
\author{Hirohisa~A.~Tanaka}\affiliation{Department of Particle Physics and Astrophysics, SLAC, Stanford University, Menlo Park, CA, 94025, USA}
\author{Wei~Wang}\affiliation{Sun Yat-Sen University, Guangzhou, China}
\author{Jaehoon~Yu\vspace{0.1in}}\affiliation{Department of Physics, University of Texas at Arlington, Arlington, TX, 76019, USA}
\author{\\ \textbf{Liaisons:} K~S.~Babu}\affiliation{Department of Physics, Oklahoma State University, Stillwater, OK, 74078, USA}
\author{Robert~H.~Bernstein}\affiliation{Muon Department, Fermi National Accelerator Laboratory, Batavia, IL, 60510, USA}
\author{Erin~Conley}\affiliation{Department of Physics, Duke University, Durham, NC, 27708, USA}
\author{Albert~De Roeck}\affiliation{EP Department, CERN, Geneva, 1211, Switzerland}
\author{Alexander~I.~Himmel}\affiliation{Neutrino Division, Fermi National Accelerator Laboratory, Batavia, IL, 60510, USA}
\author{Jay Hyun~Jo}\affiliation{Department of Physics, Yale University, New Haven, CT, 06511, USA}
\author{Claire~Lee}\affiliation{Fermi National Accelerator Laboratory, Batavia, IL, 60510, USA}
\author{Tanaz ~A. ~Mohayai}\affiliation{Neutrino Division, Fermi National Accelerator Laboratory, Batavia, IL, 60510, USA}
\author{Kim~J.~Palladino}\affiliation{Department of Physics, University of Oxford, Oxford, OX1 3RH, UK}
\author{Vishvas~Pandey}\affiliation{Neutrino Division, Fermi National Accelerator Laboratory, Batavia, IL, 60510, USA}
\author{Mayly~C.~Sanchez}\affiliation{Department of Physics, Florida State University, Tallahassee, FL, 32306, USA}
\author{Yvonne~Y.Y. ~Wong}\affiliation{School of Physics, The University of New South Wales, Sydney, NSW, 2052, Australia}
\author{Jacob~Zettlemoyer}\affiliation{Neutrino Division, Fermi National Accelerator Laboratory, Batavia, IL, 60510, USA}
\author{Xianyi ~Zhang\vspace{0.1in}}\affiliation{Lawrence Livermore National Laboratory, Livermore, CA, 94536, USA}
\author{\\ \textbf{Other Contributors:} Andrea~Pocar}\affiliation{Department of Physics, University of Massachusetts at Amherst, Amherst, MA, 01003, USA}

%% file: common/init.tex
\renewcommand{\familydefault}{\sfdefault}
\renewcommand{\thepage}{\roman{page}}
\setcounter{page}{0}

\pagestyle{plain} 

%\clearpage

\textsf{\tableofcontents}
%\clearpage

%\textsf{\listoffigures}
%\clearpage

%\textsf{\listoftables}
%\clearpage

%For acronym list to appear just after TOC, TOF, TOT
%\printnomenclature
%\clearpage

%\textsf{\listoftodos}

%\newpage
%\noindent
%\textbf{Still to be finished:} refs, final experiment table and figure, authors, acknowledgements

\renewcommand{\thepage}{\arabic{page}}
\setcounter{page}{1}

%% file: execsummary.tex
\section{Executive Summary}
\subsection{Motivation}

The discovery of neutrino mixing and oscillation, which implies that neutrinos have non-zero mass, was the first, and
remains the only, laboratory demonstration of non-Standard Model physics. Neutrino mass requires the addition of new degrees of freedom --- new interactions --- to the Standard Model Lagrangian, but we do not yet know what form that will take or the profound impact it will have on particle physics. Over the past two decades, a successful and growing neutrino physics program has demonstrated unequivocally that neutrino mixing involves all three known flavors, has provided precision measurements of many of the parameters describing three-flavor neutrino mixing, and has greatly improved our understanding of neutrino properties and their interactions with matter. However, anomalous results have raised intriguing questions about the possibility of additional particles or interactions and our most fundamental questions about neutrinos remain unanswered. What is the origin of neutrino mass and why are neutrino masses so tiny compared to other Standard Model particles? Why is neutrino mixing so different from mixing in the quark sector? Is the three-flavor neutrino picture complete? The answers to these questions will lead to extensions of the Standard Model and are fundamentally linked to our understanding of the Universe.

The physics {\it of} neutrinos and the physics that can be done {\it with} neutrinos
extends from sub-eV to EeV -- across eighteen orders of magnitude -- and touches nearly every other area of particle physics.
Right-handed neutrinos, which are arguably the most minimal addition to the Standard Model that can accommodate neutrino masses, could be a portal to a potential dark sector. Many of the ideas about how to generate neutrino mass also address questions like the baryon asymmetry of the universe, or lead to new phenomena observable at current and future colliders, or predict rare processes like the lepton-number-violating neutrinoless double beta decay. It is still experimentally allowed for neutrinos to possess a wide range of surprising properties, including relatively large couplings to new (light) states, which would lead to observable effects in experiments. Oscillation of high-energy neutrinos traveling over very long distances can provide a critical test of our understanding of space-time. Neutrino oscillation, being a manifestation of quantum interference, is uniquely sensitive to tiny effects, like the phase difference caused by neutrino mass, and thus is used to look for new interactions or to test the foundations of quantum mechanics. 

As a result of the weakness of weak interactions, neutrinos can also serve as messengers, carrying information about the physics taking place and the structures inside otherwise inaccessible systems such as the core of the Earth, the core of the Sun, supernovae, and active galactic nuclei. The information provided by neutrinos becomes even more valuable when combined with other data, e.g., from cosmic rays, gamma rays, and gravitational waves; neutrinos thus form a critical component of multi-messenger astronomy. Neutrinos can also be tools for more terrestrial concerns, such as nuclear nonproliferation.

The weakness of neutrino interactions and the smallness of their masses makes neutrino experiments particularly challenging, requiring bright sources and large, sensitive detectors. Over the past decade there has been an explosion of new techniques and technologies for producing and observing neutrinos. These technological developments will likely impact other fields directly -- there are particularly strong synergies with dark matter detection -- and they also imply that neutrino experiments are excellent tools for looking for non-neutrino physics beyond the Standard Model complementing other probes of beyond-the-standard-model (BSM) physics like collider experiments. Large neutrino detectors allow to put strong constraints on baryon-number-violating processes. Searches for very weakly coupled degrees of freedom, which may or may not have a direct connection to neutrino physics, will be performed at neutrino oscillation experiments and in dedicated experiments.  Examples of such searches use existing machines, like the LHC, or future machines, like a muon collider, that lend themselves naturally to producing unique neutrino beams. 

In the coming decade, the international high-energy physics community is committed to realizing next-generation neutrino oscillation experiments that will answer many of our remaining questions about neutrino oscillation, with a particular focus on CP violation and testing the three-flavor neutrino paradigm.  These experiments also have the potential to discover additional physics beyond the Standard Model. The nuclear physics community is poisded to build at least two ton-scale neutrinoless double beta decay experiments that will explore the parameter space favored by the inverted mass ordering under the light neutrino exchange mechanism. Conceptual designs for upgrades and next-next-generation experiments with even broader physics scope and probing deeper into parameter space are being developed. There is remarkable diversity in the programs studying neutrino properties, using neutrinos as astrophysical messengers, and supporting the neutrino physics program with phenomenology studies, computing and algorithm developments, cross-section modeling and measurements, and detector and neutrino source R\&D. The components of this broad program are all complementary and each is necessary to realize the promise neutrinos hold for deepening our understanding of the Universe.

\subsection{Science Drivers}
\label{sect:drivers}

Many of the science drivers in neutrino physics derive from questions which arise directly from the discovery of neutrino oscillation and with it the realization that neutrino masses are non-zero.  The community has made tremendous progress in addressing these questions, and as a result, the international landscape sports a rich portfolio of on-going experiments and construction of major new experiments~\cite{SajjadAthar:2021prg}.  The scientific questions motivating both the on-going program and the near-future program remain valid.  
Many fundamental questions remain about the values and nature of neutrino masses:

\begin{itemize}

\item {\bf How are the masses ordered?}  This is the question of whether  the electron neutrino mostly is made up of the heaviest mass eigenstate or not. The answer directly informs models of neutrino mass generation and has important practical implications for absolute mass-scale searches as well as neutrinoless double beta decay. The current and near future program 
may provide the first indication of the answer to this question, which will be answered definitively by DUNE.

\item {\bf What are the neutrino masses?} 
The magnitude of neutrino mass will provide important information about the scale at which it is generated. Direct kinematic mass measurements provide a limit of around 0.8~eV, with the current program expected to eventually reach a sensitivity of 0.2~eV and longer-term efforts aiming for beyond that. Limits from cosmology are currently at the 0.1-0.5 eV level and the sensitivity will reach the 0.01~eV range over the next decade. 

\item {\bf What is the origin of neutrino masses?}  We currently have no good indication of the scale at which neutrino masses are generated and there is a wide variety of models for neutrino mass generation.  These models admit many possibilities not available for the other fermions in the Standard Model. Related to this is the quest for a theory of flavor. Some of these mass and flavor models may lead to directly observable phenomena; some may not.  Thus, a combination of circumstantial evidence provided by precision measurements and rare process searches across frontiers may aid in testing these models. Many models of flavor lead to correlations between observables and these correlations can be tested with precision measurements.

\item {\bf Are neutrinos their own antiparticles?} Without additional symmetries or interactions, neutrinos are likely to have a Majorana mass term, which would be unique among the fermions. A direct experimental consequence of a Majorana mass term would be neutrinoless double beta decay. The current experimental program is sensitive to half-lives of order $10^{26}$\,years and the future ton-scale program could potentially probe the entire region allowed in the case of inverted mass ordering.

\end{itemize}

Beyond the immediate questions arising from the existence of neutrino mass, the study of neutrino properties allows us to address other major open issues:

\begin{itemize}

\item {\bf Do neutrinos and antineutrinos oscillate differently?}  The fact that the Universe consists of matter and not antimatter hints at a fundamental asymmetry between matter and antimatter, in the form of CP violation. We know that the known sources of CP violation via quark mixing and from the QCD Lagrangian are insufficient to account for the composition of the Universe. Neutrino oscillation is sensitive to one new CP phase and if neutrinos are Majorana particles, there are two additional neutrino-sector CP phases which may contribute to matter-antimatter asymmetry.

\item {\bf Is the three-flavor picture of neutrino mixing complete?} Precision measurements of neutrino oscillation in different channels and at different energies and baselines can be compared with each other and with predictions of three-flavor oscillation probabilities to determine whether the picture of three-flavor oscillation is complete, or whether there is as-yet undiscovered physics impacting neutrino mixing and oscillation. The first signs of a breakdown of the three-flavor picture could be present in a number of so-called ``anomalies," which are so-far unexplained experimental findings.
\end{itemize}

Our understanding of TeV-scale physics has seen a significant evolution with the advent of LHC results, and as a consequence the phase space for dark matter has exponentially increased. In particular, there is a renewed interest in looking for new states below the electroweak scale and for very small couplings to these new states:

\begin{itemize}

\item {\bf Discovering new particles, interactions and the unknown } Neutrino properties are not as well constrained as those of the other known particles and thus admit new interactions at a strength comparable to their Standard Model interaction strength. They may also have unexpected properties. This in turn makes neutrinos a uniquely sensitive tool to probe the physics of a wide range of dark matter models as well as for generic searches for low-scale new physics. At the same time, neutrino experiments combine high luminosity sources of photons, nuclear and meson decays with very sensitive and large detectors.  Neutrino experiments are thus discovery-class facilities for a wide range of models of BSM physics, many of which have been conceived only  in view of recent LHC results.

\end{itemize}

Owing to current and future advances in detection capabilities, neutrinos can be used as a tool to study a wide variety of phenomena.

\begin{itemize}
\item {\bf Neutrinos as astrophysical messengers}  Neutrinos are the second most abundant known particle in the Universe -- after photons -- and play an important role in many astrophysical phenomena and cosmology. Studying neutrinos allows us to learn about a wide range of environments, like stellar fusion processes, supernova explosions, nucleosynthesis, and the origin of the highest-energy 
particles ever observed.  Cosmology is sensitive to the number of neutrinos, the sum of their masses, and to potential new neutrino interactions. In return, the more precisely we can determine neutrino properties in the laboratory, the fewer degrees of freedom there are in fits to cosmological data.

\end{itemize}

All these science drivers need a supporting program of a broad and diverse nature, including applications for societal impact.  Our current modeling of neutrino-nucleus interactions needs to be improved to fully exploit the precision physics program offered by future experiments. Improving it will require cooperation among experimentalists making new measurements; nuclear and particle theorists developing new models; computing professionals enabling the preservation and interoperability of the data, theory, and software tools. A better understanding of existing neutrino beams and novel neutrino sources also requires continued effort. Theory will clearly play a central role in neutrino science and is an integral part of the future program.

A new generation of neutrino detectors based on liquid and gaseous noble element time projection chambers presents a range of unprecedented computing challenges. These detectors produce very large data volumes whose character is distinct from collider data (very large events rather than very many events) and requires innovation in hardware and software. 
The on-going transition to a new computing paradigm with a much greater focus on heterogeneous computing resources (GPUs, FPGAs, etc.) based at high-performance computing facilities will require adaptation from experiments at all scales. However, it also presents opportunities, for example using advanced machine learning algorithms to address the challenge of reconstructing the high-resolution images produced by modern TPCs.  
Many neutrino experiments have specific performance requirements, e.g., low-energy particle detection or low background detectors, requiring detector R\&D. This R\&D has strong synergy with technologies relevant for QIS; neutrino oscillation also provides a fertile test bed for our understanding of entanglement and other concepts in QIS.

Finally, underground facilities are needed by many experiments to reduce cosmogenic background. Design and management of these facilities requires a unique combination of expertise in mining, civil engineering, and science, and often involves collaboration with communities and commercial mines.

A common thread among the widely varying issues addressed by these programs is the need to collaborate across disciplines and sectors. Commitment to community engagement, including discourse with other scientific disciplines, a diverse and welcoming environment within HEP, and education and outreach, facilitates these necessary collaborations.

\subsection{Breadth and Diversity of the Neutrino Program}

\begin{figure}[thbp]
    \centering
    \includegraphics[width=0.615\linewidth]{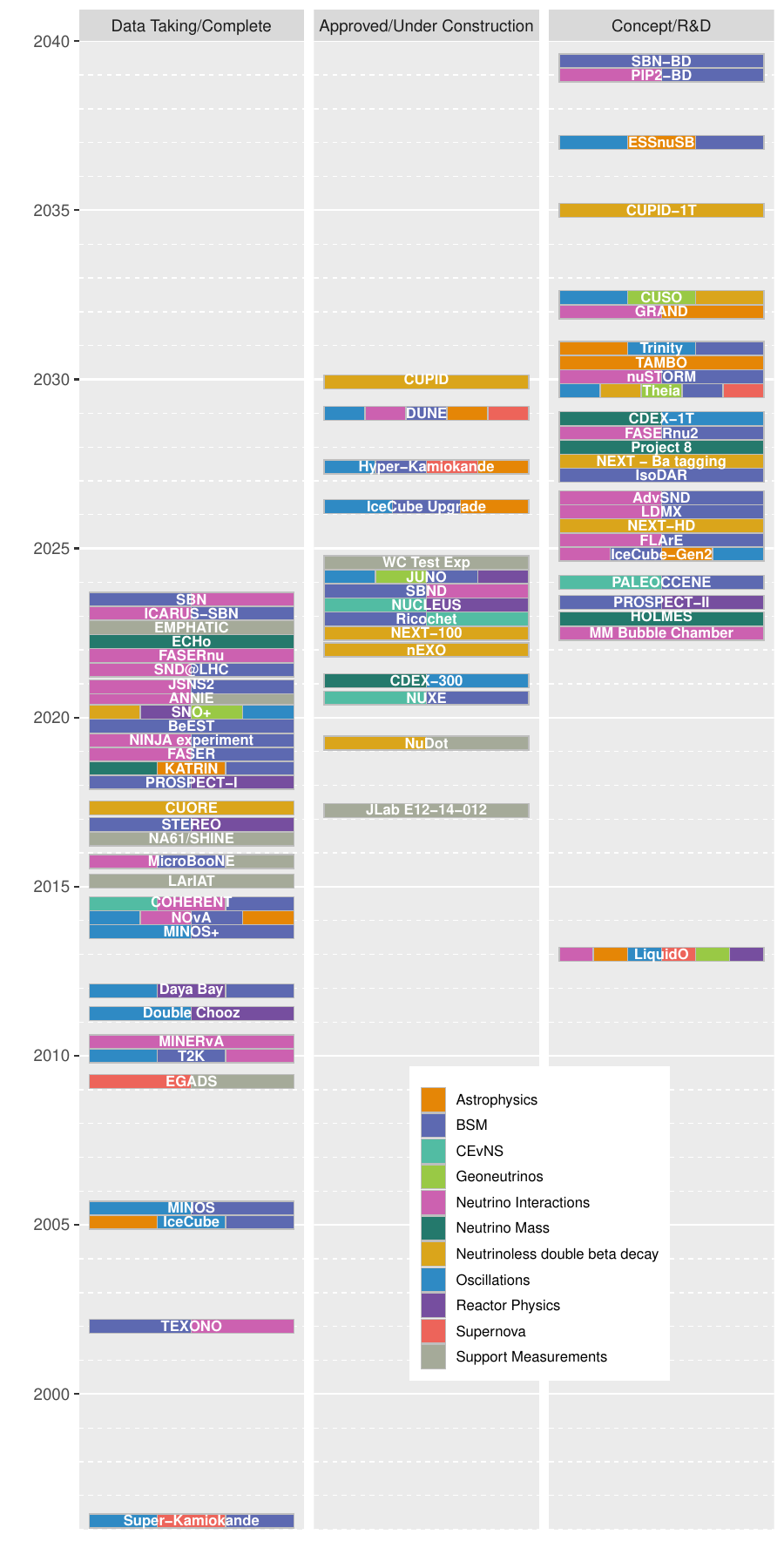}
    \caption{Summary plot of experiments indicating rough time scale and range of physics topics addressed by each.  The information is provided by the collaborations.  Details and references are found in Tab.~\ref{tab:experiments}. }.
    \label{fig:experiments}
\end{figure}

The Neutrino Frontier includes a wide range of activities spanning the full range of scales, including DUNE as a major international program with more than a thousand collaborators; medium-scale experiments such as those making up the short-baseline neutrino program; contributions to international experiments; smaller-scale experiments at accelerators, reactors, and spallation neutron sources; down to
tabletop experiments and blue-sky R\&D activities.   Many examples are described in the topical group reports and a comprehensive list of Neutrino Frontier experiments is provided in Tab.~\ref{tab:experiments}, as well as a summary graphic in Fig.~\ref{fig:experiments} created from information provided by community members.

All of these efforts have the potential to make paradigm-changing discoveries or innovations or provide necessary inputs to experiments that will make these discoveries, and they will work in synergy to address the science drivers described in Sec.~\ref{sect:drivers}. A program incorporating breadth and diversity of efforts at different scales does more than just increase the chances for major discoveries. Such a program is healthy for future scientific progress by making space for creative thinking, and also provides training opportunities to ensure a capable workforce in the long term.
The last P5 report recommended: \textit{“Maintain a program of projects of all scales, from the largest international projects to mid- and small-scale projects”}. The Neutrino Frontier endorses continued and enhanced support of this recommendation.

Cross-pollination between Frontiers and other fields of science offers further opportunities.  Physics topics within the Neutrino Frontier overlap strongly with each other, and also with other Frontiers.     
In areas where programs may have synergies, careful attention is merited to enable collaborative stewardship by traditionally separate entities.
Examples include instrumentation for dark matter and neutrinoless double beta decay searches, which connects particle astrophysics, nuclear physics, and neutrino physics; the study of neutrino-nucleus interactions, which can probe nuclear structure as well as support measurements of neutrino properties; the detection of high-energy neutrinos with far-forward experiments at the LHC; and the Cosmic Frontier programs that provide insight into neutrino properties from cosmological observables.  Also meriting attention are opportunities to make optimal use of national laboratories and international facilities for particle physics and other sciences, including those that have not historically hosted HEP experiments.

\subsection{Building on Current Success}

The U.S. has a strong ongoing neutrino program in pursuit of the science  drivers identified in Section~\ref{sect:drivers}.  Currently operating experimental programs are investigating anomalous results that may be early hints of new physics, making increasingly precise observations of neutrino oscillation using neutrinos from a variety of natural and artificial sources, characterizing neutrino properties including the ways in which they interact with matter, participating in multi-messenger astronomy, and extending the energy range of astrophysical neutrino detection into new territory. Searches for neutrino mass using both direct kinematic methods as well as cosmological fits, together with advanced experiments pursuing neutrinoless double beta decay, are also a mainstay of the current and future portfolio. This program is supported and enabled by developing sophisticated accelerator, instrumentation, computation, and analysis techniques to facilitate increased precision of neutrino measurements. 

The quest for the absolute mass of the neutrino has seen significant progress with tritium endpoint measurements.  Eventually current experiments will probe masses as small as 0.2 eV. In the near future, sensitivity into the sub-0.1 eV range will come from cosmology and studies of large-scale structure. Over the next decade, sensitivities down to 0.01\,eV are expected and thus the impact of neutrino mass on large-scale structure should be confirmed. At the same time there is active R\&D and new ideas to extend the reach of direct kinematic laboratory-based mass searches by either looking toward electron-capture isotopes or by improving the source strength and energy resolution of experiments using tritium or other beta emitters. Some of these approaches may also make progress towards the most ambitious goal of detection of the cosmic background neutrinos originating from the Big Bang.

 A number of successful efforts have led the way for next-generation experiments searching for neutrinoless double beta decay, which is the only near-term plausible way to answer the question of the Majorana vs. Dirac nature of the neutrino.  The Department of Energy Office of Nuclear Physics is stewarding the next phase, at ton-scale, which will push limits down to the region consistent with inverted mass ordering, given the assumption of light neutrino exchange.  Under this stewardship, at least two ton-scale experiments are planned, along with participation in international collaborations.  Exciting possibilities exist for farther-future experiments beyond the ton scale with sensitivity to parameters allowed under the assumption of normal mass ordering. For full exploration of this physics program, and to enable robust discovery, multiple isotopes and multiple technologies will be needed.

A major recent experimental milestone was the unequivocal detection of ultra-high energy neutrinos of astrophysical origin by IceCube and the first detection of an extra-galactic source in 2018. Neutrinos propagate in straight lines, neither affected by magnetic fields nor absorbed by dust and thus are the ideal tool to find the origin of the highest energy cosmic rays;
the data collected to date have disproven most models for the generation of the highest energy cosmic rays.  Neutrinos are a major contributor to the enterprise of multi-messenger astronomy.
Already now, the neutrinos detected by IceCube provide stringent bounds on neutrino properties and space-time structure. Other high-energy neutrino observatories in sea water and also using radio waves are being realized as well, expanding sky coverage and energy range. IceCube is currently undergoing the IceCube Upgrade which will pave the way to IceCube Generation 2, which will significantly extend the energy reach of the experiment.  Together with DUNE and Hyper-K, IceCube also will push the study of atmospheric neutrinos to new levels of sensitivity especially in the search for BSM physics.

A plethora of on-going and planned experiments will probe neutrino properties like magnetic moments, existence of sterile neutrinos across a wide range of masses and mixing angles, neutrino-nucleus interactions, couplings to light new particles and many others. Some of these experiments use traditional particle physics methods, but others use tools from nuclear physics and astrophysics. We expect that  the sheer breadth and diversity of the effort will be a continuing theme in neutrino science. Some of these experiments are motivated by a set of existing anomalies historically summarized under the theme of eV-scale sterile neutrinos. The experimental program going after these questions features accelerator-based approaches with stopped-meson, isotope decay at rest, and pion decay in flight as well as reactor neutrino experiments at a few-meter baselines. For the future, proposals for using radioactive sources produced either online or offline are being pursued, as well as novel types of particle beams based on kaon or muon decays. Since its discovery in 2017,  the study of coherent elastic neutrino nucleus  scattering (CEvNS) has become available as a tool for the study of neutrino interactions with strong technology synergies with dark matter searches. Dark matter experiments eventually also will become sensitive to solar neutrinos and other natural neutrinos via CEvNS. There are also ideas for experiments with exciting low-energy astrophysical neutrino programs which could also serve as DUNE far detectors.

Currently operating long-baseline oscillation experiments have made significant observations of \nue and \anue appearance and are making successful measurements of the parameters governing long-baseline oscillation. We anticipate that by the middle of the current decade there will be hints of answers to the question of the neutrino mass ordering and CP-violating asymmetry, although discovery-level measurements will not yet have been made. To accomplish this, significant progress has been made towards the next generation of oscillation experiments that are required to see this program through to its conclusion. JUNO will significantly improve the precision on a number of the neutrino mixing parameters and test the mass ordering with vacuum oscillation at the $3\,\sigma$ level; it is expected to come online in 2023. In the U.S., DUNE/LBNF was launched in response to the previous P5 as an international collaboration with goals of precisely measuring the parameters governing long-baseline neutrino oscillation.  Construction of DUNE is underway after the very successful operation of prototype detectors and detailed simulation studies to validate and optimize its experimental design. DUNE is expected to begin taking physics data by 2029. Significant progress has been made on PIP-II, the upgrade to Fermilab’s proton accelerator complex required for DUNE’s neutrino beam, with conventional facilities nearing completion. 
Hyper-K is also proceeding in Japan with an experimental strategy for studying long-baseline neutrino oscillation that is complementary to that of DUNE.  Hyper-K is expected to take data beginning in 2027. 

At the completion of this next-generation experimental oscillation program, we expect to have measurements of all the neutrino mixing angles and mass differences at the few percent level or better, with the measurement of several parameters made in different channels.  This will allow for a test of the unitarity of the mixing matrix, definitive determination of the neutrino mass ordering, and a measurement of the CP-violating phase with a resolution of 5-15 degrees (depending on its true value). This suite of measurements, in combination with searches for oscillation physics inconsistent with the three-neutrino paradigm, will address many of the outstanding questions related to the physics of neutrino mass. Additionally, future neutrino facilities enable a rich and diverse program of BSM physics probes. The combination of high-intensity proton beams and high-resolution detectors, along with the distance from the neutrino beam source to detectors, enable exploration of BSM physics parameter spaces highly complementary to BSM physics searches at Energy Frontier experiments.

DUNE is the largest project in the Neutrino Frontier portfolio and has been designed to definitively determine the neutrino mass ordering, measure the level of CP violation, and observe neutrino oscillation  over a wide range of neutrino energies. These measurements represent a test of the completeness of the three-neutrino paradigm. DUNE also has a broad physics program beyond three-flavor oscillation physics that includes multi-messenger astronomy and astrophysics, searches for a wide variety of BSM signatures, and precision SM measurements.

DUNE and Hyper-K have very different detector designs that provide complementary input to our understanding of neutrino physics. DUNE’s experimental design includes a baseline long enough to lift degeneracy between matter and CP-violating asymmetries, a high-power wideband neutrino beam covering a full oscillation period, liquid-argon-time-projection-chamber far-detector modules designed for detailed tracking of most final-state particles from neutrino interactions over a range of energies, and a near detector well matched to the systematics constraints needed for precision 
oscillation measurements. The Hyper-K experiment is based on a very large water Cherenkov detector, at a shorter baseline, in a high-power off-axis beam peaked at the first oscillation maximum.  Therefore Hyper-K is optimized to collect a large sample of beam neutrinos with a relatively narrow range of energies. Both strategies are expected to offer significant sensitivity to CP violation with very different systematics, baselines, and targets, which benefits our understanding of the underlying physics and will be especially powerful in disentangling the effects of potential new physics from three-flavor oscillation effects, as well as possible systematic effects. The experiments will collect highly complementary neutrino samples in the event of a core-collapse supernova, as DUNE is most sensitive to electron neutrinos via charged-current absorption while Hyper-K is most sensitive to antineutrinos via inverse beta decay on protons. The experiments have extremely broad and complementary physics portfolios that also include atmospheric and solar neutrinos, baryon number violation searches, and a vast array of searches for physics beyond the Standard Model.  There are significant opportunities for collaboration and shared efforts benefiting both DUNE and Hyper-K, including US-Japan cooperation on accelerator and targetry issues; measurements, modeling, and analysis techniques to address the challenges of neutrino interaction physics; measurements and modeling of neutrino beams; and eventually joint analysis of the data, as is currently being pursued by NOvA and T2K.

DUNE is being built in two phases. Phase I is an initial experiment configuration consisting of two liquid argon far detector modules; a suite of near detector components suitable for the initial measurement program, including a movable LArTPC and a muon spectrometer, as well as an on-axis detector; and a 1.2 MW proton beam. Moreover, Phase I includes the far site infrastructure for four detector modules and  all irreplaceable beamline components to support 2.4 MW beam operation. With the Phase I configuration, DUNE will be able to determine the neutrino mass ordering, make a measurement of the potentially CP-violating phase, and make improved measurements of other oscillation parameters, in addition to the broader program of astrophysics and BSM searches. In Phase II, to allow DUNE to achieve its full physics program, the third and fourth far detector modules will be added, the proton beam power will be upgraded to 2.4 MW, and the near detector will be 
upgraded to further constrain systematics for the precision oscillation measurement program. In the Phase II configuration, DUNE will make precision measurements that, in combination with each other and/or with world data, over-constrain the PMNS matrix to determine whether or not neutrino oscillation is well-described by the three-flavor paradigm. In the event that there are unknown particles or interactions impacting neutrino oscillation, these precision measurements will be the window for discovery.
Each of the Phase II upgrade components also provides an opportunity to broaden DUNE’s physics portfolio to address additional science drivers and to therefore involve a wider community, including potential opportunities for dark matter detection, neutrinoless double-beta decay measurements, and additional BSM searches. There is also significant community interest in other experimental efforts that may make use of Fermilab's upgraded proton accelerator complex.

DUNE is a world-class oscillation experiment hosted by the U.S. and LBNF is a unique facility  offering many additional scientific opportunities to the international community, including the possibility of additional detectors at the near and far sites. Not only are LBNF/DUNE's science capabilities intrinsically unique and broad, but it will affirm our ability to act successfully in this role for a large high-energy physics facility.  It will be important to demonstrate that the U.S. community and funding agencies can work together to create a long-term stable environment in order to enable international partners to commit to any future programs.

\subsection{Outreach \& Diversity, Equity, Inclusion}

Among connections between the Neutrino Frontier with the other Frontiers, perhaps the most pervasive and vital are the connections with the Community Engagement Frontier, for which issues are shared with the entire Snowmass community.  Given that neutrino physics represents a major component of the U.S. program, there are opportunities for leadership and special responsibility in addressing societal issues. 

It is critical for the community to address issues of diversity, equity and inclusion.  A work environment that is welcoming to all is both intrinsically desirable and conducive to scientific productivity.  A strong outreach and education component to the program is crucial for attracting the best talent to the particle physics community as well as for communicating the impact of our science to the public.  Neutrino physics facilities typically have a very significant impact on the surrounding communities, so it is important to be good citizens of them. Finally, neutrino physics has some specific practical applications of benefit to society, in particular for nuclear non-proliferation programs.

\subsection{Neutrino Community Aspirations for US-HEP Neutrino Frontier Science and Activities}

This section provides a concise high-level synthesis of input from the neutrino community regarding US-HEP planning. It is intended as input for the next P5, summarizing neutrino community aspirations for US-HEP activities in the next decade, with a view to beyond the next decade. 

\begin{itemize}

\item  Opportunities for advances in the neutrino sector are entwined with opportunities in many other sectors, spanning all of the Snowmass Frontiers and multiple scales of time, size and cost. 
\textbf{ A future program with a healthy breadth and balance of physics topics, experiment
sizes, and timescales, supported via a dedicated, deliberate, and ongoing funding
process, is highly desirable.} This process should also provide opportunities to explore and eventually resolve existing and future neutrino-related anomalies and to develop instrumentation and new beam technologies that will have a broad impact across the field. Furthermore, connections between programs should be carefully curated to optimize science output.

\item 
 
There has been tremendous progress on  oscillation physics with the current experiments and the DUNE/LBNF program since the last P5. However, the primary questions about the three-flavor paradigm remain unanswered, and the motivations for answering them, and probing new physics beyond the three-flavor paradigm, are undiminished.    \textbf{Completion of existing experiments and execution of DUNE in its full scope are critical for
addressing the NF science drivers}. Both Phase I and Phase II are part of the original DUNE
design endorsed by the last P5. DUNE Phase I will be built in the current decade and DUNE
Phase II (two additional far detector (FD) modules, a more capable near detector (ND), and use of the 2.4 MW beam power
from the FNAL accelerator upgrade) is the priority for the 2030s.

\item  Existing technologies enable the original DUNE physics program for both Phase I and Phase
II. However each piece of DUNE Phase II offers broader physics opportunities than originally
envisioned. \textbf{ To exploit these new opportunities, directed R\&D needs to be supported.}
These opportunities for DUNE Phase II should be explored with a process inclusive of the community at large.

\item Many questions in neutrino physics arise from theory and conversely neutrino experimental results raise many theory questions. A strong neutrino theory program is therefore essential to reap the full scientific benefit from the investment into new experimental facilities. Moreover, there is a significant amount of theory understanding required to correctly connect experimental observables and simulations with the underlying physics parameters. \textbf{Strong and continued support for neutrino theory is needed. }

\item Neutrinos have connections to practically all other sectors of particle physics as well as many adjacent disciplines, offering neutrino physicists the opportunity to be community leaders in issues of diversity, equity and inclusion (DEI). These opportunities must be embraced. \textbf{The Neutrino Frontier has a special responsibility to contribute to leadership for a cohesive, HEP-wide strategic plan for DEI and community engagement.}

\end{itemize}

\newpage

%% file: table.tex
Soudan II&Data Taking/Complete&1983&Atmospheric neutrino oscillations, search for proton decay and neutron-antineutron oscillations, muon astronomy&\cite{Soudan2:2003qqa}\\\\[-2ex]
Super-Kamiokande&Data Taking/Complete&1996&Atmospheric and solar neutrinos, supernova neutrino burst, diffuse supernova neutrinos, indirect dark matter, proton decay, far detector for T2K&\cite{Super-Kamiokande:2002weg}\\\\[-2ex]
TEXONO&Data Taking/Complete&2002&Magnetic moments (completed 2007),  nu-electron cross-section (completed 2011),  dark-matter searches (completed 2014),  nu-Nucleus coherent elastic scatter (since 2013),  Beyond Standard Model Physics (since 2014)&\cite{Wong:2015kgl}\\\\[-2ex]
MINOS&Data Taking/Complete&2005&Studies of neutrino oscillations and related processes&\\\\[-2ex]
EGADS&Data Taking/Complete&2009&Supernova neutrinos, Gd-H2O technology demonstrator&\cite{Marti:2019dof}\\\\[-2ex]
MINERvA&Data Taking/Complete&2010&Precision measurements of neutrino interactions on a variety of nuclei&\cite{MINERvA:2021csy}\\\\[-2ex]
T2K&Data Taking/Complete&2010&Neutrino oscillation, searches for exotic physics, and measurements of neutrino interactions&\cite{Abe:2011ks}\\\\[-2ex]
Daya Bay&Data Taking/Complete&2011&Oscillations (precision measurement of theta13 mixing angle and atmospheric mass splitting), search for sterile neutrinos, characterization of reactor antineutrino emission, searches for new physics. &\cite{DayaBay:2018yms}\\\\[-2ex]
Double Chooz&Data Taking/Complete&2011&Neutrino Oscillation (measurement of $	heta$13) and Neutrino Reactor (IBD mean cross-section per fission)&\cite{DoubleChooz:2019qbj}\\\\[-2ex]
MINOS+&Data Taking/Complete&2013&Studies of neutrino oscillations and related processes (using NOvA-tuned NuMI beam)&\cite{MINOS:2011rry}\\\\[-2ex]
NOvA&Data Taking/Complete&2014&Neutrino oscillations, in particular mass ordering , CP violation, and octant.  Steriles.  Neutrino Cross sections. Cosmic ray and astrophysical sources studies.   &\\\\[-2ex]
LArIAT&Data Taking/Complete&2015&Hadron scattering measurements on argon target&\cite{LArIAT:2019kzd}\\\\[-2ex]
MicroBooNE&Data Taking/Complete&2015&Sterile neutrinos, BSM searches, neutrino interaction measurements, demonstration of liquid argon technology&\cite{MicroBooNE:2016pwy}\\\\[-2ex]
The MAJORANA DEMONSTRATOR&Data Taking/Complete&2015&Neutrinoless Double-Beta Decay, Dark Matter Searches&\\\\[-2ex]
STEREO&Data Taking/Complete&2016&Sterile neutrino, 235U fission spectrum&\cite{STEREO:2019ztb}\\\\[-2ex]
NA61/SHINE&Data Taking/Complete&2016&Hadron production measurements for neutrino beams&\cite{NA61:2014lfx}\\\\[-2ex]
CUORE&Data Taking/Complete&2017&Neutrinoless double-beta decay&\cite{CUORE:2021mvw}\\\\[-2ex]
PROSPECT-I&Data Taking/Complete&2018&Sterile neutrino (BSM), Reactor Flux and Spectrum, Boosted Dark Matter, Detector Technology Development&\cite{PROSPECT:2020sxr}\\\\[-2ex]
BeEST&Data Taking/Complete&2019&KeV Neutrino Search&\cite{PhysRevLett.126.021803}\\\\[-2ex]
NINJA experiment&Data Taking/Complete&2019&Neutrino cross-section measurement, sterile neutrino search&\cite{Fukuda:2017clt,NINJA:2020gbg,NINJA:2020bvx,Suzuki:2021euh,Odagawa:2022crm}\\\\[-2ex]
KATRIN&Data Taking/Complete&2019&Neutrino mass, sterile neutrino, relic neutrinos, BSM&\cite{KATRIN:2022ayy}\\\\[-2ex]
JSNS2&Data Taking/Complete&2020&Sterile neutrino, neutrino xsec, exotic searches&\cite{JSNS2:2021hyk}\\\\[-2ex]
SNO+&Data Taking/Complete&2020&Double beta decay, reactor and geo antineutrinos, solar neutrinos&\cite{SNO:2021xpa}\\\\[-2ex]
SND@LHC - Scattering and Neutrino Detector at the LHC&Data Taking/Complete&2021&TeV neutrino interactions, lepton flavor universality, heavy flavor production in the forward region &\cite{Ahdida:2750060}\\\\[-2ex]
e4nu&Data Taking/Complete&2022&Oscillation &\cite{Ankowski:2022thw}\\\\[-2ex]
ICARUS at SBN&Data Taking/Complete&2022&Sterile neutrino, dark matter, neutrino cross section, other BSM&\cite{MicroBooNE:2015bmn}\\\\
IceCube Neutrino Observatory&Approved/Under construction&2005&High energy neutrino astrophysics. Neutrino oscillations. Sterile neutrinos. Indirect search for Dark Matter. &\cite{ACHTERBERG2006155}\\\\[-2ex]
JLab E12-14-012&Approved/Under construction&2017&Measurement of inclusive and exclusive cross sections for argon and titanium. Extraction of the spectral functions from the exclusive data.&\cite{Benhar:2014nca}\\\\[-2ex]
FASER&Approved/Under construction&2019&Cross sections at TeV energies, lepton universality, forward charm production&\cite{FASER:2019dxq,FASER:2021mtu}\\\\[-2ex]
NuDot&Approved/Under construction&2019&Double-beta Decay; Liquid scintillator R\&D&\cite{klein2022future}\\\\[-2ex]
LEGEND-200&Approved/Under construction&2022&Neutrinoless double-beta decay; dark matter searches; other BSM physics&\cite{LEGEND_LOI}\\\\[-2ex]
NEXT-100&Approved/Under construction&2022&Neutrinoless double beta decay&\cite{martin2016sensitivity}\\\\[-2ex]
SBN - Short-Baseline Neutrino Program at Fermilab&Approved/Under construction&2022&Sterile neutrinos, neutrino interactions, BSM searches&\cite{MicroBooNE:2015bmn}\\\\[-2ex]
nEXO&Approved/Under construction&2022&Neutrino-less Double Beta Decay&\cite{Adhikari_2021}\\\\[-2ex]
NUCLEUS&Approved/Under construction&2023&CEvNS at reactors, non-proliferation&\cite{Strauss_2017}\\\\[-2ex]
Jianmen Underground Neutrino Observatory (JUNO)&Approved/Under construction&2023&Neutrino mass ordering, precision measurement of three oscillation parameters, solar neutrinos, geoneutrinos, atmospheric neutrinos, measurement of reactor antineutrino spectrum, proton decay, searches for BSM physics (possibly extending to neutrinoless double-beta decay in a second phase)&\cite{CONNIE:2022hna}\\\\[-2ex]
SBND&Approved/Under construction&2023&Sterile neutrinos, neutrino interactions, BSM searches&\cite{Machado:2019oxb}\\\\[-2ex]
Ricochet&Approved/Under construction&2023&Neutrino coherent scattering from reactor neutrinos.  Beyond standard model searches&\cite{Abdullah:2022zue}\\\\[-2ex]
Water Cherenkov Test Experiment&Approved/Under construction&2024&Pion scattering, secondary neutron production, water Cherenkov detector response&\cite{Barbi:2712416}\\\\[-2ex]
IceCube Upgrade&Approved/Under construction&2026&Neutrino Oscillations, Indirect dark matter search, astrophysics&\cite{Ishihara:2019aao}\\\\[-2ex]
Hyper-Kamiokande&Approved/Under construction&2027&Neutrino oscillations (CP violation, mass hierarchy etc), Nucleon Decays, Neutrino Astrophysics&\cite{Hyper-Kamiokande:2018ofw}\\\\[-2ex]
DUNE&Approved/Under construction&2029&CP violation, mass ordering, precision oscillation physics, Supernovae, solar neutrinos, sterile neutrinos, precision tests of 3-flavor paradigm, nucleon decay, direct DM detection, other BSM searches&\cite{DUNE:2022aul}\\\\[-2ex]
LiquidO&Concept/R\&D&2013&Technology envisaged for neutrino physics in general above 1MeV.&\cite{Cabrera:2019kxi}\\\\[-2ex]
COHERENT&Concept/R\&D&2014&Neutrino scattering, BSM searches, accelerator-produced dark matter, neutrino electromagnetic moments,  sterile oscillations, nuclear structure&\cite{Akimov:2022oyb}\\\\[-2ex]
CDEX-300&Concept/R\&D&2020&Neutrino mass, CP violation&\cite{CDEX_DBD_program}\\\\[-2ex]
NUXE&Concept/R\&D&2020&Reactor neutrino CEvNS, neutrino magnetic moment, sterile neutrino&\cite{universe7030054}\\\\[-2ex]
ANNIE&Concept/R\&D&2020&Neutrino interactions, detector R\&D&\cite{ANNIE:2017nng}\\\\[-2ex]
EMPHATIC&Concept/R\&D&2022&Neutrino flux &\\\\[-2ex]
ECHo&Concept/R\&D&2022&Neutrino mass&\cite{Kovac:2022aea}\\\\[-2ex]
HOLMES&Concept/R\&D&2022&Neutrino mass&\cite{Nucciotti:2018vyc}\\\\[-2ex]
Modern Modular Bubble Chamber LDRD&Concept/R\&D&2022&Broad terms are to take measurements of neutrino cross sections.&\cite{alvarezrusoetal:2022}\\\\
FASERnu&Concept/R\&D&2022&TeV neutrino interactions&\cite{Feng:2022inv,Anchordoqui:2021ghd,FASER:2019dxq}\\\\[-2ex]
PROSPECT-II&Concept/R\&D&2023&Sterile neutrino (BSM), Reactor Flux and Spectrum, Boosted Dark Matter, Detector Technology Development&\cite{Andriamirado:2022psq}\\\\[-2ex]
LEGEND-1000&Concept/R\&D&2024&Neutrinoless double-beta decay; dark matter searches; other BSM physics&\cite{LEGEND_LOI}\\\\[-2ex]
PALEOCCENE&Concept/R\&D&2024&CEvNS at reactors, dark matter, non-proliferation&\cite{Cogswell:2021qlq,Alfonso:2022meh}\\\\[-2ex]
IceCube-Gen2&Concept/R\&D&2025&Neutrino astronomy, supernovae, cosmic rays, atmospheric neutrinos, neutrino oscillations, sterile neutrinos, neutrino interactions&\cite{IceCube-Gen2:2020qha}\\\\[-2ex]
SBC-CEvNS&Concept/R\&D&2025&Reactor CEvNS&\cite{flores2021physics}\\\\[-2ex]
FLArE&Concept/R\&D&2026&TeV neutrino cross-section, lepton flavor universality, heavy flavor production in the forward region, dark matter direct detection, milli-charged particles&\cite{Feng:2022inv,Anchordoqui:2021ghd}\\\\[-2ex]
LDMX&Concept/R\&D&2026&EN scattering, dark matter searches&\cite{Akesson:2022vza}\\\\[-2ex]
FASERnu2&Concept/R\&D&2026&TeV neutrino cross-section, tau neutrinos, lepton flavor universality, heavy flavor production in the forward region&\cite{Anchordoqui:2021ghd}\\\\[-2ex]
AdvSND - Advanced Scattering and Neutrino Detector at the LHC&Concept/R\&D&2026&TeV neutrino cross-section, lepton flavor universality, heavy flavor production in the forward region &\cite{Feng:2022inv,Anchordoqui:2021ghd}\\\\[-2ex]
NEXT-HD&Concept/R\&D&2026&Neutrinoless double beta decay&\cite{adams2021sensitivity}\\\\[-2ex]
IsoDAR&Concept/R\&D&2027&Sterile neutrino, weak mixing angle, non-standard neutrino interactions, light X particles&\cite{isodar_physics_2022,isodar_web}\\\\[-2ex]
Project 8&Concept/R\&D&2028&Neutrino mass scale with aimed sensitivity of 40 meV/c$^2$.&\cite{Project8:2022wqh}\\\\[-2ex]
CDEX-1T&Concept/R\&D&2028&Neutrino mass, CP violation&\cite{CDEX_DBD_program}\\\\[-2ex]
NEXT with barium tagging&Concept/R\&D&2028&Neutrinoless double beta decay&\cite{mcdonald2018demonstration}\\\\[-2ex]
CUPID&Concept/R\&D&2029&Neutrinoless double-beta decay&\cite{CUPID:2019imh}\\\\[-2ex]
TAMBO: Tau Air-Shower Mountain-Based Observatory&Concept/R\&D&2030.&Make precision measurements of high-energy astrophysical tau neutrinos.&\cite{Romero-Wolf:2020pzh}\\\\[-2ex]
Trinity&Concept/R\&D&2030&Astrophysical neutrinos, oscillations, bsm&\cite{https://doi.org/10.48550/arxiv.2109.03125}\\\\[-2ex]
nuSTORM&Concept/R\&D&2030&Neutrino interactions an cross sections, sterile neutrinos, and BSM searches&\cite{Ahdida:2020whw}\\\\[-2ex]
Theia&Concept/R\&D&2030&CP violation, neutrinoless double beta decay, CNO solar neutrinos, 8B solar neutrino spectrum, precision geoneutrino flux, nucleon decay, supernova neutrinos, DSNB, sterile neutrinos&\cite{Askins:2019oqj}\\\\[-2ex]
CUSO (Case Underground Salt Observatory)&Concept/R\&D&2032&Solar neutrinos, geoneutrinos, 0vbb&\cite{https://doi.org/10.48550/arxiv.2203.06262}\\\\[-2ex]
The Giant Radio Array for Neutrino Detection (GRAND)&Concept/R\&D&2032&Ultra high energy neutrino detection, ultra high energy neutrino cross sections, astrophysics&\cite{GRAND:2018iaj}\\\\[-2ex]
CUPID-1T&Concept/R\&D&2035&Neutrinoless double-beta decay&\cite{CUPID:2022wpt}\\\\[-2.1ex]
European Spallation neutrino Super Beam (ESSnuSB)&Concept/R\&D&2037&CP violation, cosmological neutrinos, sterile neutrinos, proton lifetime&\cite{Alekou:2022emd}\\\\[-2.1ex]
SBN-BD (SBN Beam Dump Experiment)&Concept/R\&D&2039&Accelerator-produced dark matter, sterile neutrino&\cite{Toups:2022knq}\\\\[-2.1ex]
PIP2-BD (PIP-II Beam Dump Experiment)&Concept/R\&D&2039&Accelerator-produced dark matter, ALPs, testing short-baseline neutrino anomalies, precision tests of the SM, neutrino-nucleus cross sections&\cite{Toups:2022yxs}

%% file: common/final.tex
\cleardoublepage
\printglossaries

\bibliographystyle{utphys}
\bibliography{common/citations,tablerefs,tdr-citedb,whitepapers}